\documentclass[fleqn,usenatbib]{mnras}

\usepackage{newtxtext,newtxmath}

\usepackage[T1]{fontenc}
\usepackage{ae,aecompl}
\usepackage[usenames]{color}
\definecolor{purple}{RGB}{160,32,240}

\bibstyle{mnras}

\usepackage{subfig}  
\usepackage{lastpage}  
\usepackage{graphicx}	
\usepackage{amsmath}	
\usepackage{amssymb}	
\usepackage{soul}

\title[Scatter in the Stellar Mass--Halo Mass Relation]{Constraining Scatter in the Stellar Mass--Halo Mass Relation for Haloes Less Massive than the Milky Way}

\author[M.\ Allen, et al.]{
Magdelena Allen,$^{1}$
Peter Behroozi,$^{2}$
and Chung-Pei Ma$^{1,3}$
\\
$^{1}$Department of Astronomy, University of California, Berkeley, CA 94720, USA\\
$^{2}$Department of Astronomy and Steward Observatory, University of Arizona, Tucson, AZ 85721, USA\\
$^{3}$Department of Physics, University of California, Berkeley, CA 94720, USA
}

\newcommand{\Msun}{\mathrm{M}_\odot}

\pubyear{2018}

\usepackage{etoolbox}
\makeatletter
\patchcmd\@combinedblfloats{\box\@outputbox}{\unvbox\@outputbox}{}{%
  \errmessage{\noexpand\@combinedblfloats could not be patched}%
}%
\makeatother

\begin{document}
\label{firstpage}
\pagerange{\pageref{firstpage}--\pageref{lastpage}}
\maketitle

\begin{abstract}
Most galaxies are hosted by massive, invisible dark matter haloes, yet little is known about the scatter in the stellar mass--halo mass relation for galaxies with host halo masses $M_{h}\le 10^{11}\Msun$. Using mock catalogues based on dark matter simulations, we find that two observable signatures are sensitive to scatter in the stellar mass--halo mass relation even at these mass scales; i.e., conditional stellar mass functions and velocity distribution functions for neighbouring galaxies.  We compute these observables for 179,373 galaxies in the Sloan Digital Sky Survey (SDSS) with stellar masses $M_\ast > 10^9 \Msun$ and redshifts 0.01 $< z <$ 0.307.  We then compare to mock observations generated from the \textit{Bolshoi-Planck} dark matter simulation for stellar mass--halo mass scatters ranging from 0 to 0.6 dex.  The observed results are consistent with simulated results for most values of scatter ($<$0.6 dex), and SDSS statistics are insufficient to provide firm constraints. However, this method could provide much tighter constraints on stellar mass--halo mass scatter in the future if applied to larger data sets, especially the anticipated Dark Energy Spectroscopic Instrument Bright Galaxy Survey. Constraining the value of scatter could have important implications for galaxy formation and evolution. 

\end{abstract}



\section{Introduction}

In the $\Lambda$CDM paradigm, galaxies grow at the centres of virialized, self-bound dark matter haloes.  Halo formation is hierarchical, such that smaller self-bound \textit{satellite} haloes can be found within the virial radii of larger haloes;  haloes that are not contained within a larger virialized structure are known as \textit{central} haloes.

While galaxy stellar mass correlates with halo mass, this correlation is not perfect.  At halo masses $M_h>10^{12}\Msun$, there are multiple ways to estimate scatter in the galaxy--halo connection, including galaxy clustering, group catalogues, direct X-ray masses, and satellite kinematics.  These methods have converged on $0.15-0.23$ dex of stellar mass scatter for such haloes, with no apparent dependence on halo mass \citep{More09,Reddick13,Tinker17,Kravtsov18}.

Considerably less is known about the stellar mass scatter for lower-mass haloes.  The shape of the stellar mass--halo mass relation results in galaxy formation becoming rapidly more inefficient for haloes with masses lower than $M_h\sim 10^{12}\Msun$ \citep{Moster10,Moster13,Behroozi10,Behroozi13,Garrison-Kimmel14}.  Thus, low-mass galaxies have fewer satellites (limiting group catalogue and satellite kinematics approaches), no mass-dependence in their bias \citep[limiting clustering techniques]{Tinker10}, and their surrounding gas is too cold to emit detectable levels of X-rays.

At the same time, there has been increased interest in the stellar mass--halo mass scatter for $M_h<10^{12}\Msun$ due to the ``too big to fail'' problem \citep{BK11}, wherein dark matter-only simulations overpredict the numbers of dense satellites.  One way to resolve this problem is for low-mass ($M_h \lesssim 3\times 10^9\Msun$) satellites to have large amounts of scatter in stellar mass at fixed halo mass, which is indeed a generic prediction from hydrodynamical simulations \citep{Sawala16,Munshi17}.  If, on the other hand, the scatter remains tight, then several authors have proposed that warm or self-interacting dark matter models are necessary to resolve the problem \citep{BK11,Elbert15,Garrison-Kimmel17,GC17}.

Here, we describe a method to measure scatter in lower-mass haloes that is based on forward modelling.  Briefly, we use abundance matching \citep{Nagai05,Conroy06,Behroozi10,Moster10,Reddick13} to populate haloes with galaxies in a dark matter simulation with different amounts of scatter.  In this technique, galaxies in a given observed volume are assigned by rank order in mass to dark matter haloes (also rank ordered by mass) in an equivalent simulated volume; these assignments are then perturbed iteratively until the desired scatter is achieved.  We show that velocity distribution functions and conditional stellar mass functions are both sensitive to the level of stellar mass--halo mass scatter.  Intuitively, larger scatters allow lower-mass haloes to host larger galaxies, hence reducing satellite counts in conditional stellar mass functions.  At the same time, smaller galaxies can be hosted by larger-mass haloes, thus broadening velocity distributions.

The use of two different scatter-sensitive techniques is important, as a key uncertainty is when satellite haloes (i.e., haloes within the virial radius of a larger halo) are considered merged.  Generally, higher-resolution simulations track satellite haloes longer \citep{Onions12}, and so would give higher predictions for conditional stellar mass functions \textit{and} velocity distribution functions at fixed scatter.  The same is true if satellites are tracked after disappearance using ``orphan'' techniques \citep[e.g.,][]{Kitzbichler08}.  Hence, using two different techniques allows for self-consistently breaking this degeneracy.

We present the observational and simulated data sets in \S \ref{sec:data}, our method for calculating velocity distribution functions and conditional stellar mass functions in \S \ref{sec:methods}, results and discussion in  \S \ref{sec:results}, and conclusions in \S \ref{sec:conclusions}.
The analysis here adopts a flat, $\Lambda$CDM cosmology ($\Omega_M = 0.307$, $\Omega_\Lambda=0.693$, $h=0.678$, $n_s = 0.96$, $\sigma_8 = 0.823$) consistent with the \textit{Planck} 2015 results \citep{Planck15}.  Stellar masses assume a \cite{Chabrier03} initial mass function, and halo masses use the virial spherical overdensity definition from \cite{Bryan98}.


\section{Data} \label{sec:data}

\subsection{Observations} \label{sec:obs}

Here, we describe the selection of 179,373 Sloan Digital Sky Survey target galaxies with $M_\ast >10^9 \Msun$ and 0.01 $< z <$ 0.307 (\S \ref{sec:obs_cat}), the corrections applied (\S \ref{sec:obs_corr}), and the method of error calculation (\S \ref{sec:obs_err}).

\subsubsection{Catalogue} \label{sec:obs_cat}

Redshifts are taken from the Sloan Digital Sky Survey (SDSS) Release 10 \citep{Ahn2014}, and have over 90\% completeness for galaxies brighter than the SDSS r-band apparent magnitude limit of 17.77.  As determined in \S3 of \cite{Behroozi15}, this corresponds to a stellar mass completeness limit as a function of redshift given by: 
\begin{equation} \label{eq:r_SDSS}
17.77 = r < -0.25 - 1.9\log_{10}\Big(\frac{M_*}{M_{\odot}}\Big) + 5\log_{10}\Big(\frac{D_L(z)}{10pc}\Big)
\end{equation}
where $M_*$ is the stellar mass and $D_L(z)$ is the luminosity distance for our assumed cosmology. Median total stellar masses and star formation rates (SFRs) are from the MPA-JHU value-added catalogue \citep{Kauffmann03,Brinchmann04}, updated for the imaging and spectroscopy in the SDSS Data Release 7 \citep{Abazajian09} and both calculated assuming \cite{Chabrier03} initial mass functions (IMFs). All galaxy targets are taken with $z>0.01$ to minimize the effect of peculiar velocities on their inferred distances.  

\begin{figure}
	\includegraphics[width=\columnwidth]{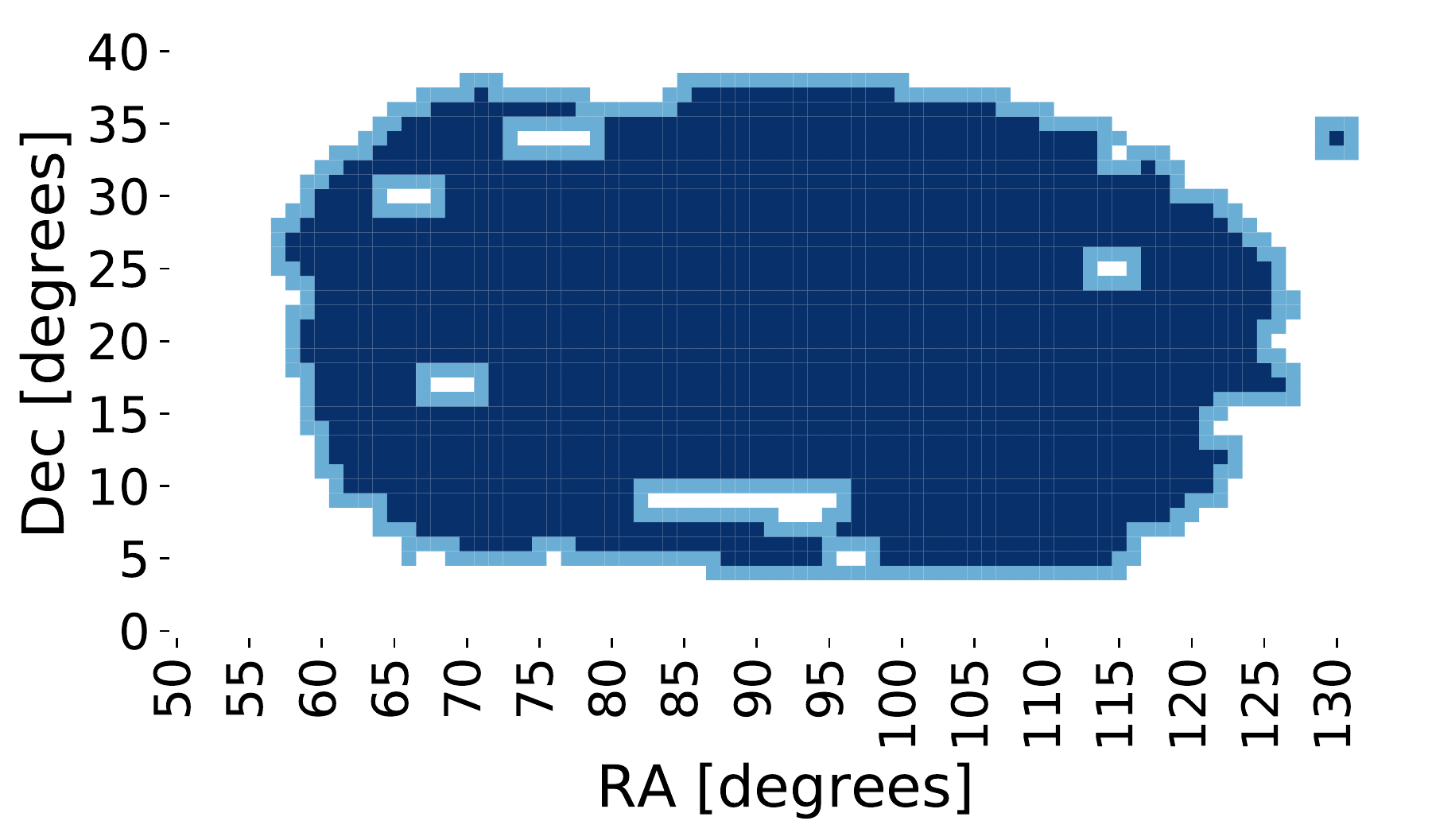}
    \caption{Map showing SDSS survey regions included (dark blue) and excluded (light blue) for this analysis.}
    \label{fig:skymap}
\end{figure}

As described in \S \ref{sec:iso_cuts}, we use isolation criteria to preferentially select central galaxies for this analysis.  The fraction of isolated galaxies was found to vary significantly in regions close to survey boundaries.  Hence, we excluded galaxies within any bin (2 degrees in right ascension by 2 degrees declination) bordering survey edges (Figure \ref{fig:skymap}) from our isolated samples, but allowed such galaxies to be included in total neighbour counts. The final cut included 179,373 targets with $M_\ast >10^9 \Msun$ over 5706 square degrees of sky, with a maximum observed redshift of 0.307 and median observed redshift of 0.076. 

\subsubsection{Galaxy Weights} \label{sec:obs_corr}

Close galaxy pairs (separated by less than 55 arcseconds) are under-selected in the SDSS due to fibre collisions, so we apply a statistical weight $w_C$ to each galaxy in close pairs to compensate. This correction is first estimated as a function of angular separation, based on a functional model of the incompleteness of spectroscopic pairs compared to photometric pairs as defined in \citealt{Patton02}, \S 5.3. We take $w_C$ to be 3.08, as calculated for a similar SDSS sample in \citealt{Patton13}, \S 3. 

Due to the SDSS magnitude limit given in Equation \ref{eq:r_SDSS}, a galaxy of any given stellar mass is only detectable to a certain maximum redshift distance $z_{max}$. We therefore assign each galaxy a volume correction weight 
\begin{equation} \label{eq:w_V}
w_V = 1/V(z_{max})
\end{equation}
where $V(z)$ is the comoving volume out to redshift $z$. To correct for under-selected central galaxies, we use these weights when averaging neighbour counts for the distributions detailed in \S \ref{sec:csmf} and \S \ref{sec:v_dist}. 

However, even if a central galaxy is detected at $z_1$, its neighbours may still be underrepresented, and so a different correction is made:
\begin{equation}
w_V = \frac{1}{\max(V_c,V_n)}
\end{equation}
where $V_c$ is the maximum observable volume of the central galaxy and $V_n$ is the same for the neighbour; $w_V$ is thus the inverse of the maximum observable volume for the pair. The final weight applied to each galaxy is then $w_C w_V$.

\subsubsection{Error Analysis} \label{sec:obs_err}

Errors on neighbour and central galaxy counts (due to sample variance) are calculated via spatial bootstrap resampling. The observed catalogue is divided into regions of 10x10 degrees (in RA and Dec) and randomly resampled 100 times to produce reconstructed sky surveys within <1\% of the original area. Regions of 10x10 degrees are chosen to preserve local structure and resample on scales where the Universe becomes homogeneous (>10 Mpc/h), though not much difference in error estimates is seen between using 2x2 and 10x10 degree regions in the resampling. The distributions described in \S \ref{sec:methods} are then computed for the resampled catalogues and the standard deviation calculated.  Our simulated catalogues are much larger in volume than our observed sample, and so we assume that the error budget is dominated by  observational sample variance.

\subsection{Simulations} \label{sec:sims}

Here, we describe the generation of simulated dark matter haloes (\S \ref{sec:sim_gen}), the addition of orphan satellites (\S \ref{sec:sim_orphans}), and the process of assigning galaxy masses to account for variable scatter in the halo mass--stellar mass relation (\S \ref{sec:sim_masses}). 

\subsubsection{Dark Matter Simulation} \label{sec:sim_gen}

We base our mock catalogues (one for each tested value of scatter) on the \textit{Bolshoi-Planck} dark matter simulation \citep{Klypin14,RP16}.  The simulation followed 2048$^3$ particles ($\sim 8$ billion) each of mass $1.55\times 10^{8}\Msun/h$ in a periodic box of comoving side length 250 Mpc$/h$ from $z=100$ to $z=0$, using the \textsc{art} code \citep{Kravtsov97,Kravtsov99}.  The adopted flat $\Lambda$CDM cosmology was consistent with \textit{Planck} 2015 results ($\Omega_M = 0.307$, $\Omega_\Lambda=0.693$, $h=0.678$, $n_s = 0.96$, $\sigma_8 = 0.823$).  Haloes were found using the \textsc{Rockstar} phase-space halo finder \citep{Rockstar} and the \textsc{Consistent Trees} merger tree code \citep{BehrooziTree}. 

\subsubsection{Orphan Satellites} \label{sec:sim_orphans}

A significant uncertainty with simulations is how long satellite haloes persist before disruption.  In large cosmological simulations, it is often necessary to include ``orphan'' satellites to match galaxy clustering \citep{Kitzbichler08,Moster18,Behroozi18}; here, we generate catalogues with orphans as they are required to best match observations.

Orphan satellite haloes were added to the \textit{Bolshoi-Planck} halo catalogues following the prescription in \cite{Behroozi18}.  Briefly, satellites that disappear in the simulation are presumed to continue orbiting their last host halo; we integrate their continued motion using a softened gravity law:
\begin{equation}
\dot{\mathbf{v}} = \frac{GM_\mathrm{host}(<r)}{(r+0.1R_\mathrm{vir,host})^2}
\end{equation}
where $M_\mathrm{host}(<r)$ is the total mass (including dark matter and baryons) enclosed within the satellite distance $r$, $R_\mathrm{vir,host}$ is the virial radius of the host halo, and the softening is performed to avoid unphysical hard scattering between the satellite and the host halo. The choice to include orphan haloes is motivated by comparison with observations in Appendix \ref{a:orphans}. 

Note that the effects of dynamical friction are not accounted for in this model, because those effects are most significant when the two interacting objects are of similar mass. Here, the vast majority of orphan satellites that disrupt in the simulation are of much smaller mass than their host haloes, and therefore the effects of dynamical friction are minimal. However, it is important to note that these effects could share parameter space with orphan lifetimes, and we do not have any good constraints on these effects. 

Satellite mass loss follows \cite{Jiang16}, with the modification that satellites do not lose mass on infalling orbits and lose mass at twice the rate on outgoing orbits \citep{Behroozi18}.  Satellites are assumed to disrupt (merge with central halo) once their maximum circular velocity ($v_\mathrm{max} \equiv \max_R \sqrt{GM(<R)/R}$) falls below $0.6 v_\mathrm{Mpeak}$, where $v_\mathrm{Mpeak}$ is the maximum circular velocity at the time the halo reached peak mass.  Identical merging criteria are applied to orphan and ordinary satellites: the stellar mass of the satellite is merged into the central galaxy if it is within $\sim0.25R_\mathrm{vir,host}$ and added to the diffuse halo otherwise. This prescription results in $\sim 25\%$ more satellites independent of mass, and was found in \cite{Behroozi18} to give the best match to galaxy autocorrelation functions.

\begin{figure}
    \centering
    \includegraphics[width=\columnwidth]{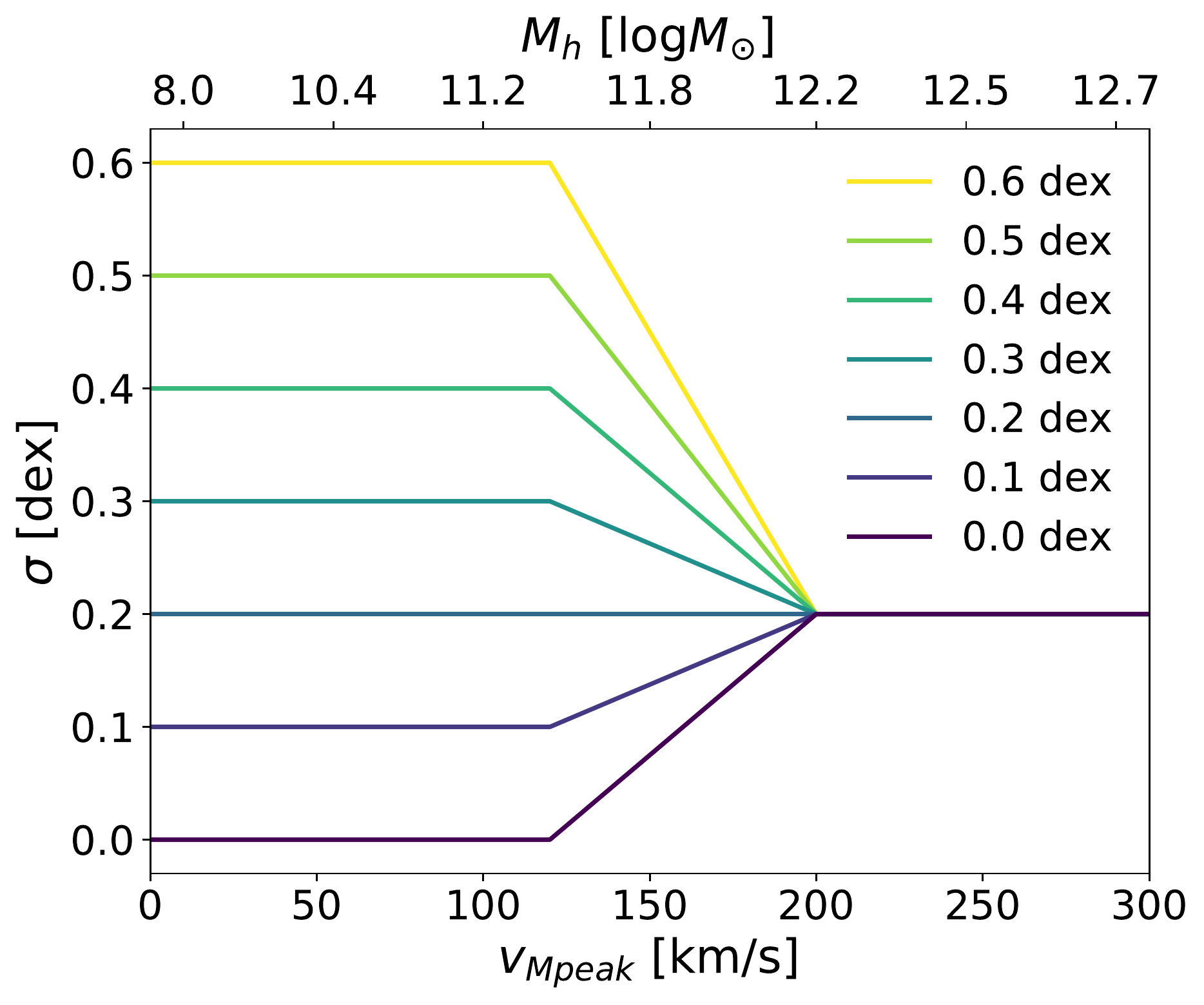}
    \caption{Functional form of scatter in dex for each tested model, as a function of the maximum circular velocity at the time the halo reached peak mass ($v_{\mathrm{Mpeak}}$).  The equivalent peak halo mass appears on the top axis.  All models are consistent with prior observable constraints; i.e., that the scatter in stellar mass is $\sim 0.2$ dex for halos of mass $M_h > 10^{12}\Msun$.}
    \label{fig:scatter_models}
\end{figure}

\subsubsection{Assigning Galaxy Masses} \label{sec:sim_masses}

Galaxy masses are then assigned using an abundance matching approach \citep[e.g.,][]{Marinoni02,Nagai05,Conroy06,Reddick13} to the dark matter halo catalogue (including orphans) at $z=0$.  We compute galaxy number densities from the selected SDSS regions (included in Figure \ref{fig:skymap}, totaling 5706 square degrees) using the inverse volume weights described in Eq. \ref{eq:w_V}.  We compute the resulting number of galaxies as a function of mass that would be expected within the simulation volume.  These galaxies are ordered by decreasing mass and assigned to haloes in order of decreasing $v_\mathrm{Mpeak}$ with zero scatter.  We then introduce $v_\mathrm{Mpeak}$-dependent log-normal scatter of $\sigma(v_\mathrm{Mpeak})$ via the following iterative algorithm:
\begin{enumerate}
\item Each scattered mass $M_\ast'$ is drawn from a log-normal distribution centred on the asssigned stellar mass $M_\ast$ and of width $\sigma(v_\mathrm{Mpeak})$. This introduces the correct scatter, but deforms the shape of the total stellar mass function.
\item Redo abundance matching (with the original SDSS galaxy number densities) to haloes ordered by decreasing $M_\ast'$.  This corrects the shape of the stellar mass function, but alters the distribution of the scatter.
\item Recompute the median $M_\ast(v_\mathrm{Mpeak})$ relation from the stellar masses assigned in step (ii).  End algorithm if median relation changes by less than 1\% from previous iteration.  This step approximates the change in the median $M_\ast(v_\mathrm{Mpeak})$ relation necessary to reproduce the correct stellar mass function after adding scatter.
\item Assign stellar masses to halos with zero scatter according to the computed median $M_\ast(v_\mathrm{Mpeak})$ relation and go to step (i).
\end{enumerate}
This can be recognized as a simple deconvolution algorithm, as discussed further in \cite{Behroozi10}.

Because constraints on the scatter for massive haloes ($M_h > 10^{12}\Msun$) are tighter than for less massive haloes, at $\sim 0.2$ dex \citep{Reddick13}, we adopt the following functional form for $\sigma(v_\mathrm{Mpeak})$, with a given scatter $X$ for low-mass haloes:
\begin{align}
\frac{\sigma(v_\mathrm{Mpeak},X)}{\mathrm{dex}} = \begin{cases}
0.2 & \textrm{if }v_\mathrm{Mpeak}>200\, \textrm{km s}^{-1}\\
X & \textrm{if }v_\mathrm{Mpeak}<120\, \textrm{km s}^{-1}\\
X+(0.2-X)L(v_\mathrm{Mpeak})& \textrm{otherwise}
\end{cases}
\end{align}
where $L(v_\mathrm{Mpeak})$ linearly interpolates from 0 to 1 as $v_\mathrm{Mpeak}$ goes from 120 to 200 km s$^{-1}$:
\begin{equation}
    L(v_\mathrm{Mpeak}) = \frac{v_\mathrm{Mpeak}-120\,\textrm{km s}^{-1}}{200\,\textrm{km s}^{-1}-120\,\textrm{km s}^{-1}} 
\end{equation}
This forms a bridge between the well-constrained scatter of high-mass haloes and the values tested for the unknown scatter of low-mass haloes. The exact bounds on $v_\mathrm{Mpeak}$ are an arbitrary modelling choice, and smoothly vary the scatter from 0.2 for $\sim{}10^{12}\Msun$ haloes to $X$ for $\sim 10^{11}\Msun$ and smaller haloes. Here, we test seven different values of $X$, i.e., 0.0, 0.1, 0.2, 0.3, 0.4, 0.5, and 0.6; the resulting functions for $\sigma(v_\mathrm{Mpeak})$ are plotted in Fig. \ref{fig:scatter_models}.

\section{Methods} \label{sec:methods}

Here, we describe methods for selecting isolated galaxies (\S \ref{sec:iso_cuts}) as well as for measuring conditional stellar mass functions (CSMFs; \S \ref{sec:csmf}) and velocity distribution functions (VDFs; \S \ref{sec:v_dist}) of neighbouring galaxies around our selected galaxy samples.  The same methods are applied both to the observations from the SDSS (\S \ref{sec:obs}) and to seven mock catalogues generated from dark matter simulations (\S \ref{sec:sims}).  The mock catalogues each include a different stellar mass--halo mass scatter from  0-0.6 dex for haloes $M_h < 10^{11}\Msun$; to match observed constraints on scatter for higher-mass haloes, the input scatter is fixed at 0.2 dex for $M_h \gtrsim 10^{12}\Msun$ and smoothly varied for $10^{11}\Msun<M_h<10^{12}\Msun$ haloes (see Fig. \ref{fig:scatter_models} and \S \ref{sec:sims}). CSMFs and VDFs are calculated separately for low-mass galaxies ($10^9$-$10^{10} \Msun$) and high-mass galaxies ($10^{10}$-$10^{11} \Msun$). 

\subsection{Isolation cuts} \label{sec:iso_cuts}

To constrain the scatter in the relationship between halo mass and stellar mass for central galaxies, we first apply an isolation cut to galaxies in both simulations and observations to preferentially select central galaxies. We define purity as the fraction of selected galaxies that are true central galaxies; we define completeness as the fraction of true central galaxies that pass our cut. As discussed in Appendix \ref{a:purity_completeness}, we take cuts to maximize completeness while retaining above 90\% purity for the two galaxy mass ranges analyzed (Table \ref{tab:purity_completeness}).  For the lower-mass sample ($10^9$-$10^{10}\Msun$), we select galaxies that are the most massive within 0.5 Mpc/h projected comoving distance $r_p$ and a redshift distance $\Delta v$ of $\pm 500$ km s$^{-1}$ (5 Mpc $h^{-1}$ in comoving line of sight distance).  For the higher-mass sample ($10^{10}$-$10^{11} \Msun$), we select galaxies that are the most massive within 1 Mpc $h^{-1}$ projected comoving distance and a redshift distance of $\pm 1000$ km s$^{-1}$ (10 Mpc $h^{-1}$ in comoving line of sight distance). Using comoving distance instead of physical distance ensures a fairer comparison between the observations (at a range of redshifts; see \S \ref{sec:obs}) and simulated catalogues (at a single redshift, $z=0$; see \S \ref{sec:sim_masses}). 

\begin{table*}
\centering
\begin{tabular}{c|c|c|c|c|c}
\hline
Scatter (dex) & Mass cut ($\log{M_{\odot}}$) & $r_p$ cut (Mpc/h) & $\Delta v$ cut (Mpc/h) & Purity & Completeness \\
\hline
0.0 & 9-10 & 0.5 & 5 & 0.987 & 0.533 \\
0.6 & 9-10 & 0.5 & 5 & 0.953 & 0.513 \\
0.0 & 10-11 & 1 & 10 & 0.982 & 0.433 \\
0.6 & 10-11 & 1 & 10 & 0.912 & 0.407 \\
\hline
\end{tabular}
\caption{Purity and completeness of central galaxies for the chosen galaxy isolation cuts, as measured in our simulated galaxy catalogues.} \label{tab:purity_completeness}
\end{table*}

In the simulated galaxy catalogues, we use the distant observer approximation, so that projected distances are calculated along the $X$ and $Y$ axes, and redshift distances are calculated as $Z + \frac{v_Z}{H(z)}$, where $v_Z$ is the $Z$-velocity and $H(z)$ is the Hubble expansion rate.  

In the observations, we calculate the projected comoving distance as:
\begin{equation} \label{eq:proj_dist}
r_p = \theta_\mathrm{sep}D(z_2)
\end{equation}
where $\theta_\mathrm{sep}$ is the angular distance between the two galaxies, and $D(z_2)$ is the comoving line-of-sight distance calculated from the redshift of the neighbouring galaxy.  We use $z_2$ instead of the more common average of $z_2$ and $z_1$ (the redshift of the central galaxy) to minimize the difference between the volume searched around observed and simulated central galaxies due to the distant observer approximation.

\subsection{Conditional stellar mass function} \label{sec:csmf}


Here, we define the conditional stellar mass function (CSMF) as the number counts of neighbouring galaxies as a function of stellar mass within a cylindrical aperture and $\pm$1.5 Mpc/h redshift distance ($|\Delta v| <$ 150 km s$^{-1}$), averaged across all central galaxies.  Neighbours are required to be galaxies within the same mass range as the sample (low-mass or high-mass) being considered.  The observed CSMFs are calculated with the corrections for fibre collisions and maximum enclosed volume as described in \S \ref{sec:obs_corr}. 

For both low- and high-mass galaxies, we count neighbours within a projected distance cut of their central galaxy to maximize statistics and minimize any disagreements between the simulations and observations. This projected distance $r_p$ is comoving, as defined in Equation \ref{eq:proj_dist}. 

\begin{figure*}
    \centering
    \subfloat[Low-mass galaxies ($10^{9}$-$10^{10} \Msun$).]{
      \includegraphics[width=0.9\columnwidth]{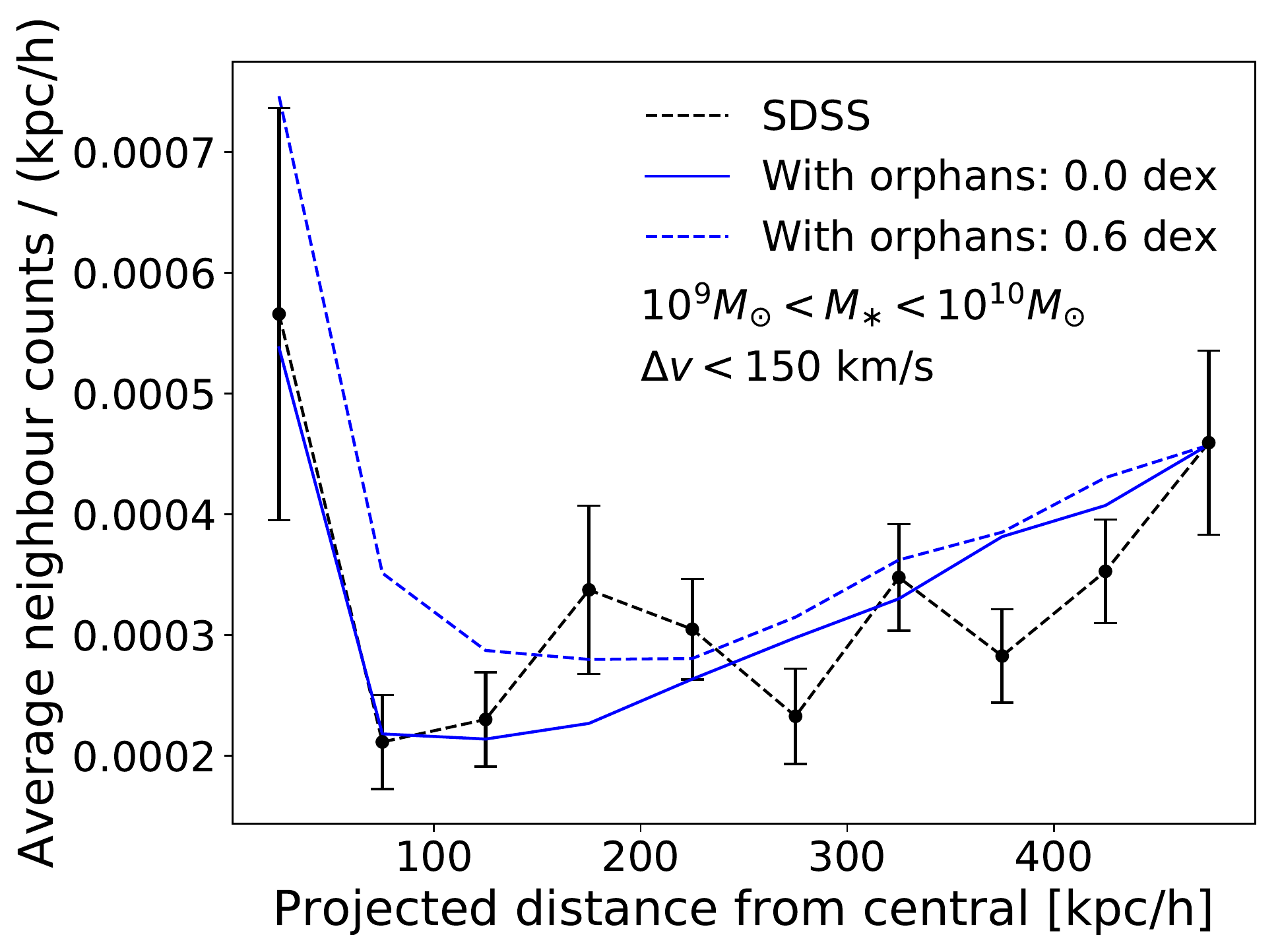}
    }\qquad
    \subfloat[High-mass galaxies ($10^{10}$-$10^{11} \Msun$).]{
     \includegraphics[width=0.9\columnwidth]{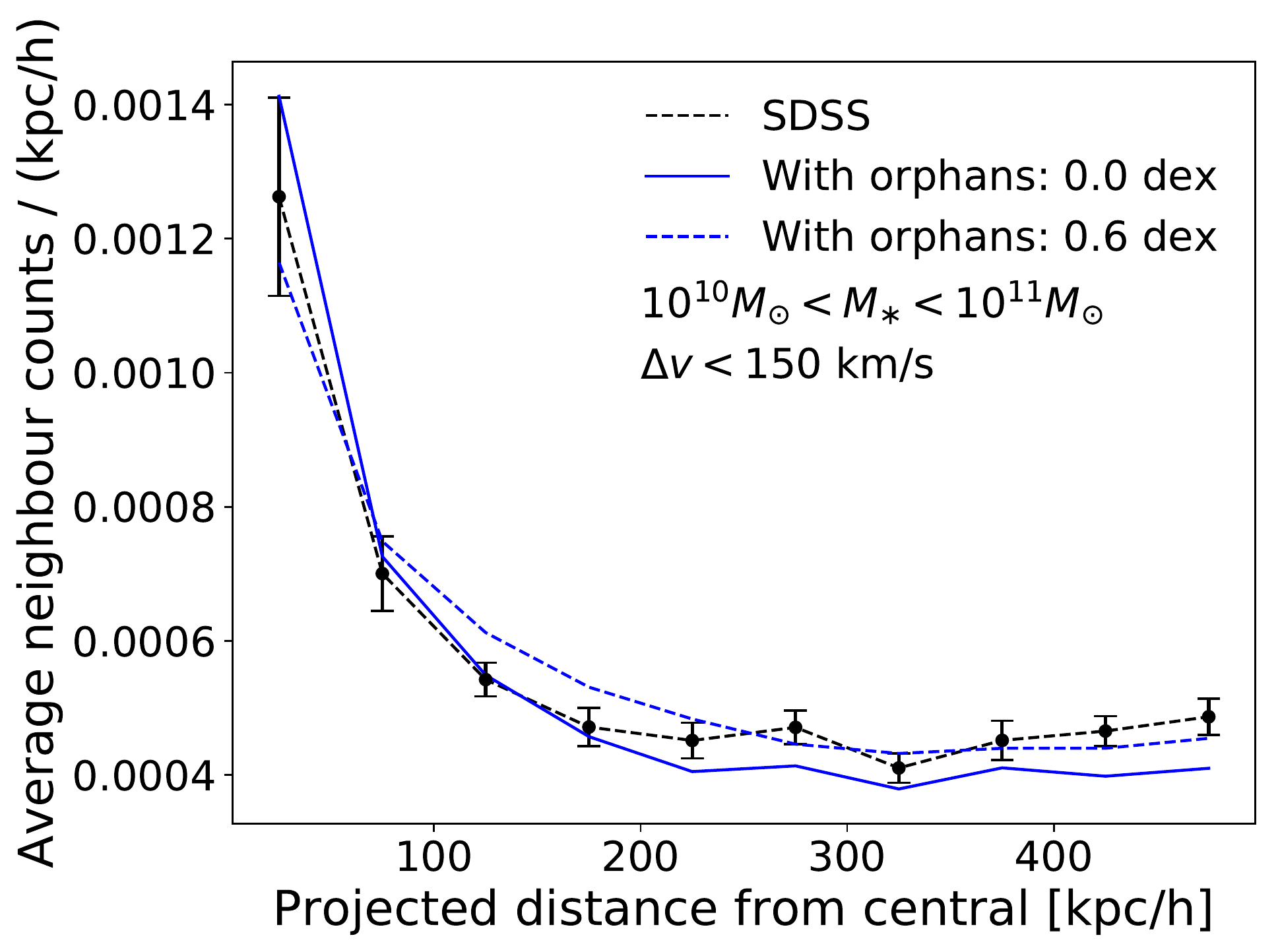}
    }
    \caption{Average neighbour counts as a function of projected distance from low-mass and high-mass galaxies.  Neighbours are galaxies in the same mass range within $\pm$1.5 Mpc h$^{-1}$ in redshift distance ($|\Delta v| <$ 150 km s$^{-1}$). The black line shows the observed distribution and the blue lines show the simulated distributions for different input scatters between stellar mass and halo mass.}
    \label{fig:n_by_r}
\end{figure*}

Average neighbour counts as a function of projected distance for low-mass and high-mass central galaxies (Figure \ref{fig:n_by_r}) show consistency of observed to simulated catalogues, and reveal that satellite galaxies are concentrated at small radii.  As expected, the low-mass galaxies are much more sensitive to the adopted scatter model.  Counts are seen to be increasing with projected distance $\gtrsim 250$ kpc/h in the low-mass sample, and this is likely an artifact of the isolation cut: selecting low-mass centrals necessarily means selecting galaxies in environments that are under-dense within the chosen projected distance cut. (This phenomena would not apply to high-mass centrals, and this trend is not observed for the high-mass sample.) As motivated in Appendix \ref{a:rp_cuts}, we take neighbours 50 $<r_p<$ 200 kpc/h for low-mass galaxies and neighbours 100 $<r_p<$ 300 kpc/h for high-mass galaxies for the highest consistency between simulated and observed catalogues, and for the best statistics.

\begin{figure}
	\includegraphics[width=\columnwidth]{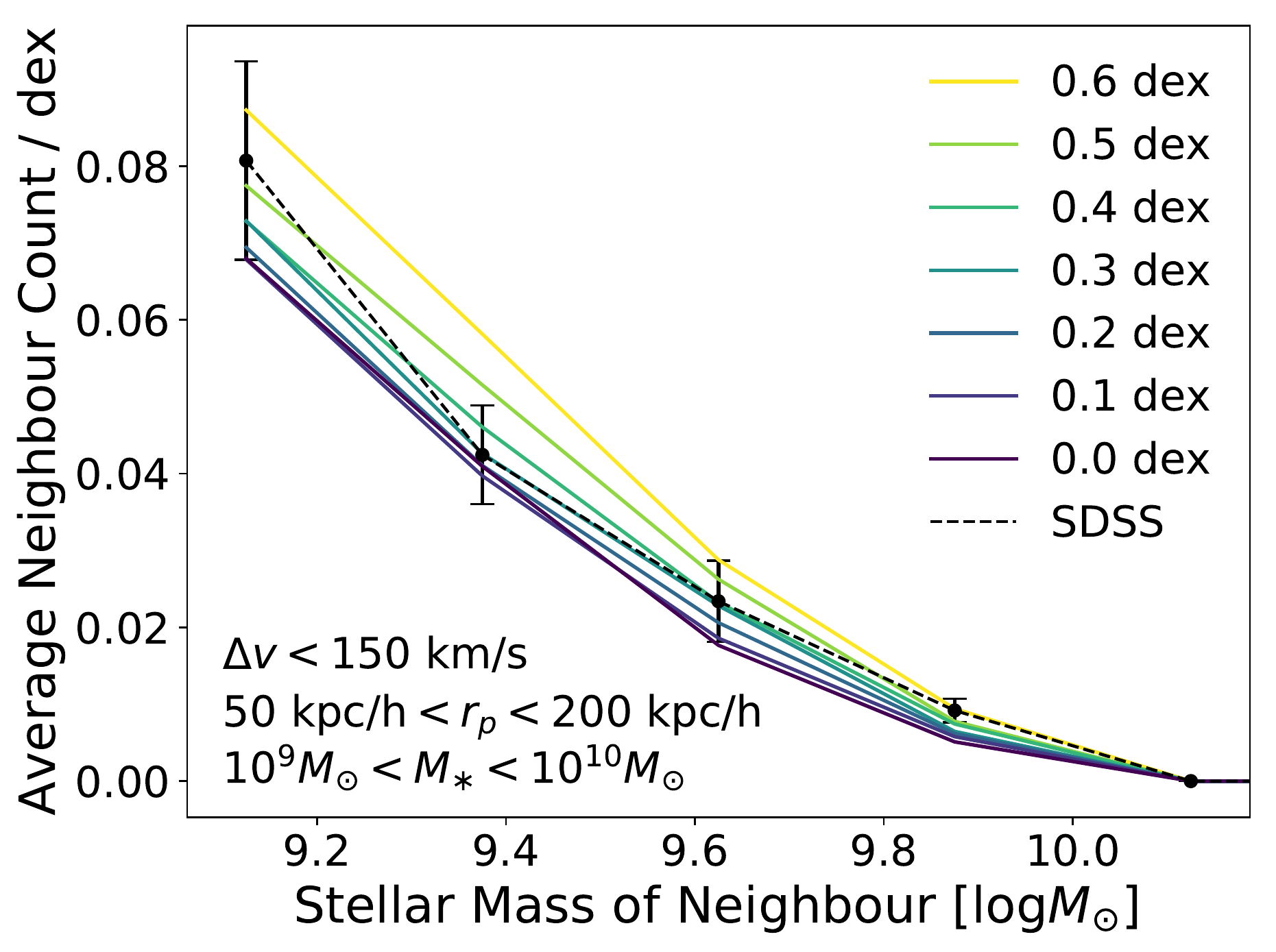}
    \caption{Comparison of observed and simulated conditional stellar mass functions for low-mass galaxies. Neighbours are galaxies in the same mass range and within 50-200 kpc/h projected and $\pm$1.5 Mpc/h redshift distance ($|\Delta v| <$ 150 km s$^{-1}$).}
    \label{fig:low_csmf}
\end{figure}

\begin{figure}
	\includegraphics[width=\columnwidth]{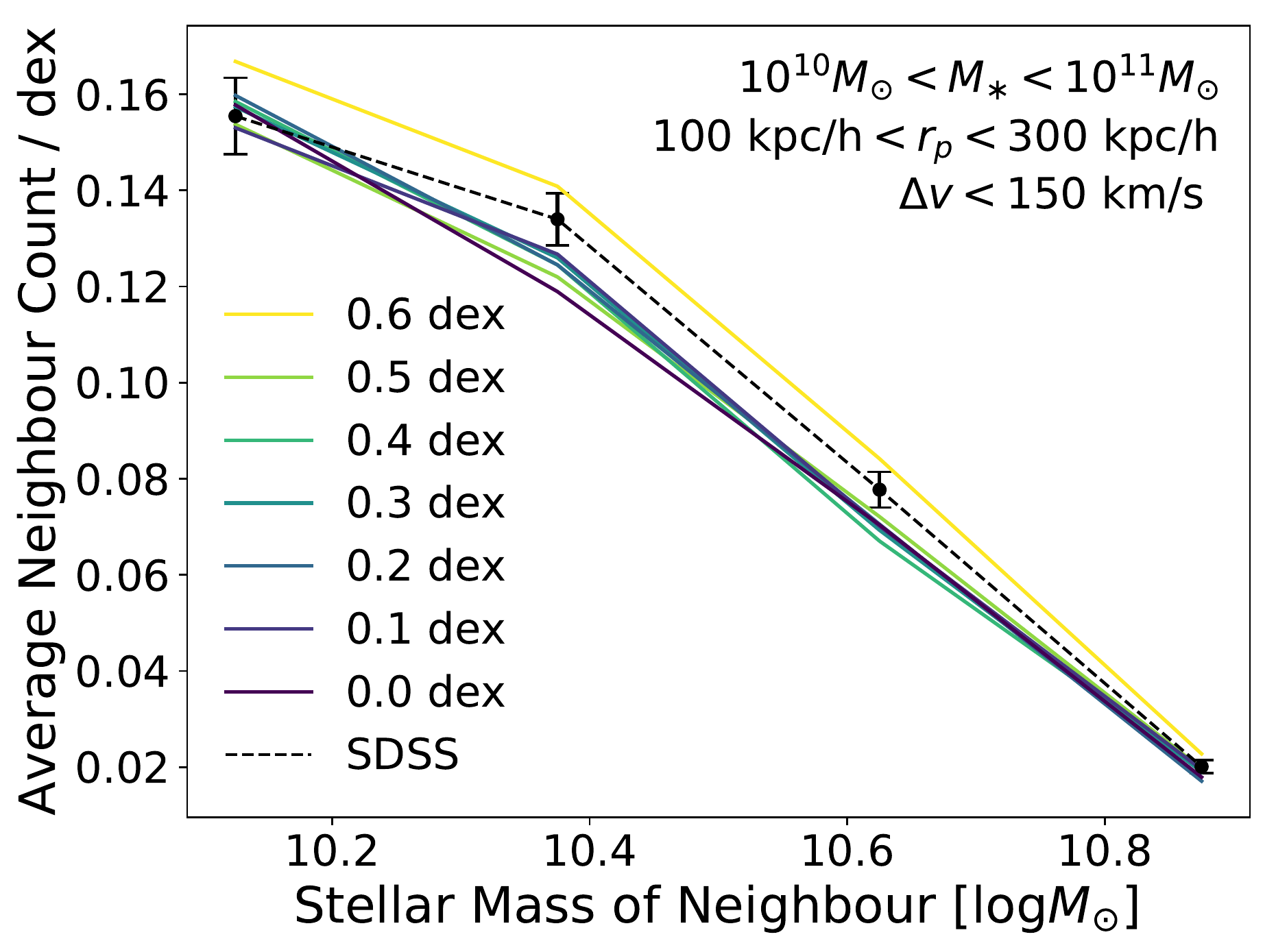}
    \caption{Comparison of observed and simulated conditional stellar mass functions for high-mass galaxies. Neighbours are galaxies in the same mass range and within 100-300 kpc/h projected and $\pm$1.5 Mpc/h redshift distance ($|\Delta v| <$ 150 km s$^{-1}$).}
    \label{fig:high_csmf}
\end{figure}

\begin{figure}
	\includegraphics[width=\columnwidth]{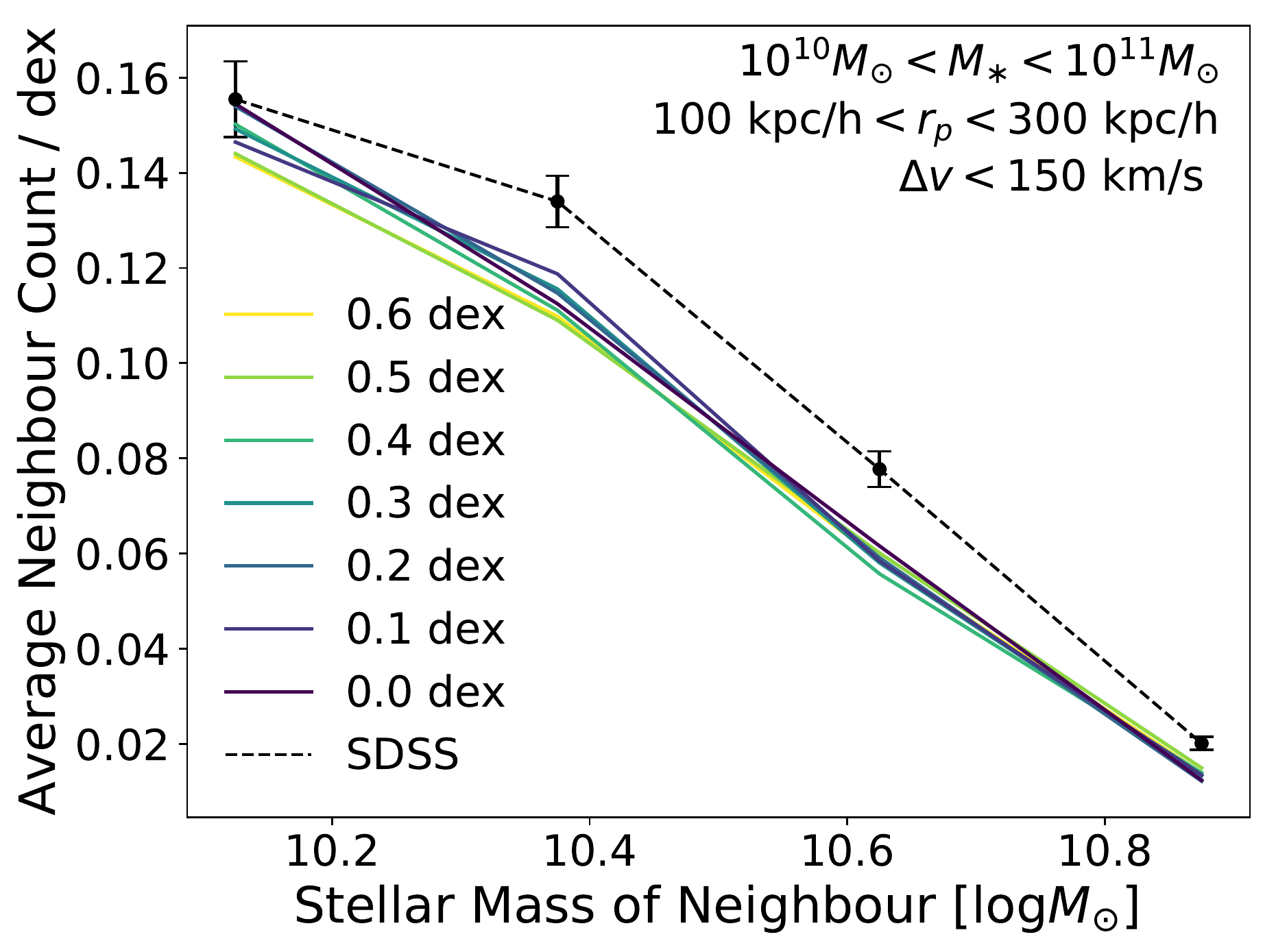}
    \caption{Comparison of observed and simulated conditional stellar mass functions for high-mass galaxies, with falsely identified centrals removed from the simulated distributions. Neighbours are galaxies in the same mass range and within 100-300 kpc/h projected and $\pm$1.5 Mpc/h redshift distance ($|\Delta v| <$ 150 km s$^{-1}$).}
    \label{fig:high_csmf_UPID}
\end{figure}

High values of scatter allow larger haloes to host smaller galaxies and smaller haloes to host larger galaxies, so the CSMFs of low-mass galaxies are expected to increase and the CSMFs of high-mass galaxies are expected to decrease. Nonetheless, because the magnitude of the scatter for high-mass haloes is fixed, the effect on the CSMF will be much larger for low-mass than for high-mass galaxies.  Low-mass galaxies clearly produce the expected trend (Figure \ref{fig:low_csmf}). High-mass galaxies show less distinction between scatters (Figure \ref{fig:high_csmf}), which is expected, but the 0.6 dex CSMF clearly deviates from the expected trend. This indicates the threshold at which scatter becomes high enough to contaminate our sample of selected central galaxies (i.e., to reduce purity). If falsely identified centrals are removed from the high-mass simulated distributions, overall counts are reduced and the sample shows decreasing neighbour counts with increasing scatter, as predicted (Figure \ref{fig:high_csmf_UPID}). 


\subsection{Velocity distribution} \label{sec:v_dist}

The velocity distribution function (VDF) gives the average number of neighbours as a function of redshift distance within a circular aperture.  Again, we require neighbours to belong to the same mass range as the sample of central galaxies considered and take projected distance cuts of 50-200 kpc/h for low-mass galaxies and 100-300 kpc/h for high-mass galaxies. Observed VDFs are calculated with the corrections described in \S \ref{sec:obs_corr}.  

\begin{figure}
	\includegraphics[width=\columnwidth]{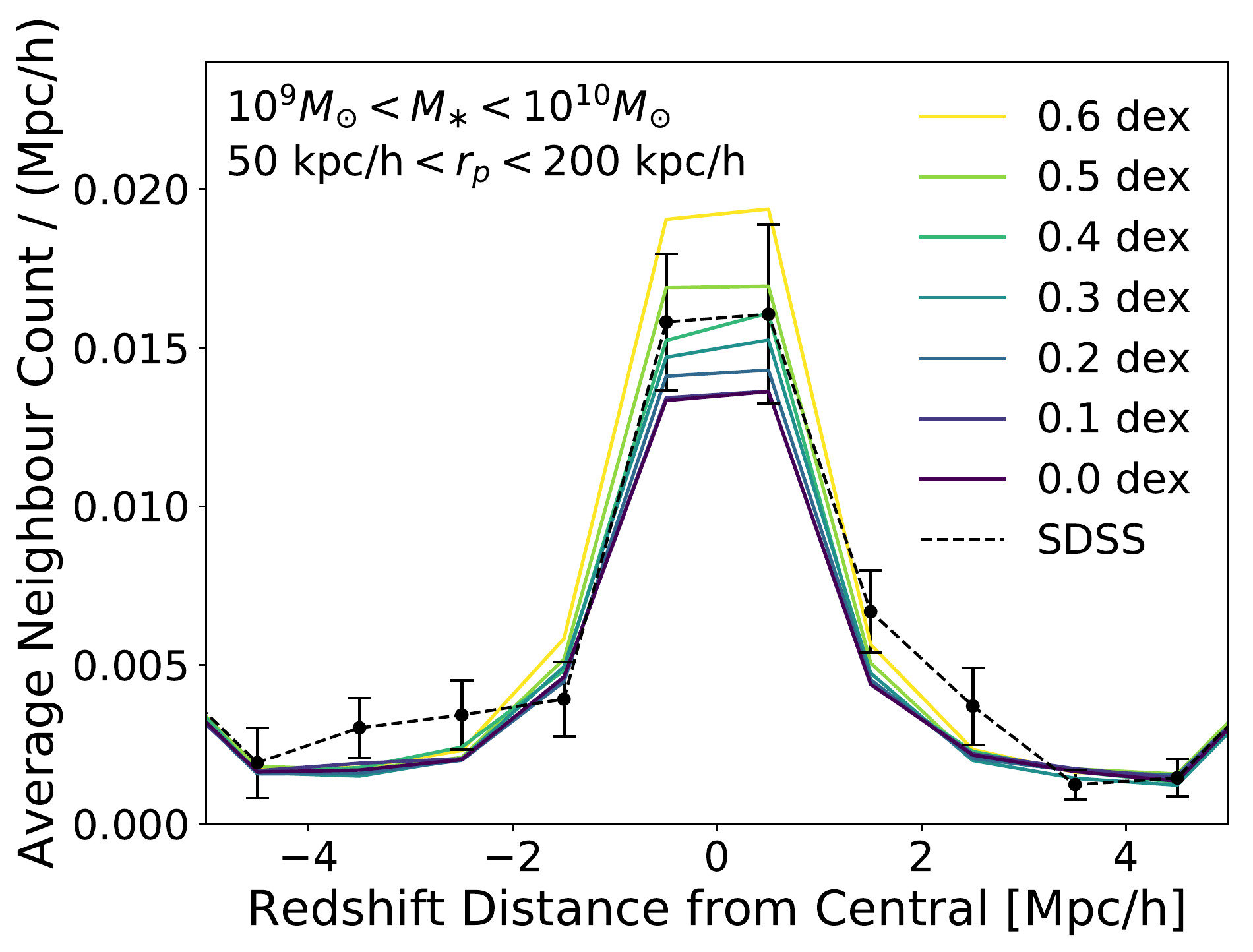}
    \caption{Comparison of observed and simulated velocity distribution functions for low-mass galaxies. Neighbours are galaxies in the same mass range and within 50-200 kpc/h projected distance. The upturn at large redshift distances is expected due to the isolation criterion (no larger galaxy within a redshift distance of 5 Mpc/h).}
    \label{fig:low_vdf}
\end{figure}

\begin{figure}
	\includegraphics[width=\columnwidth]{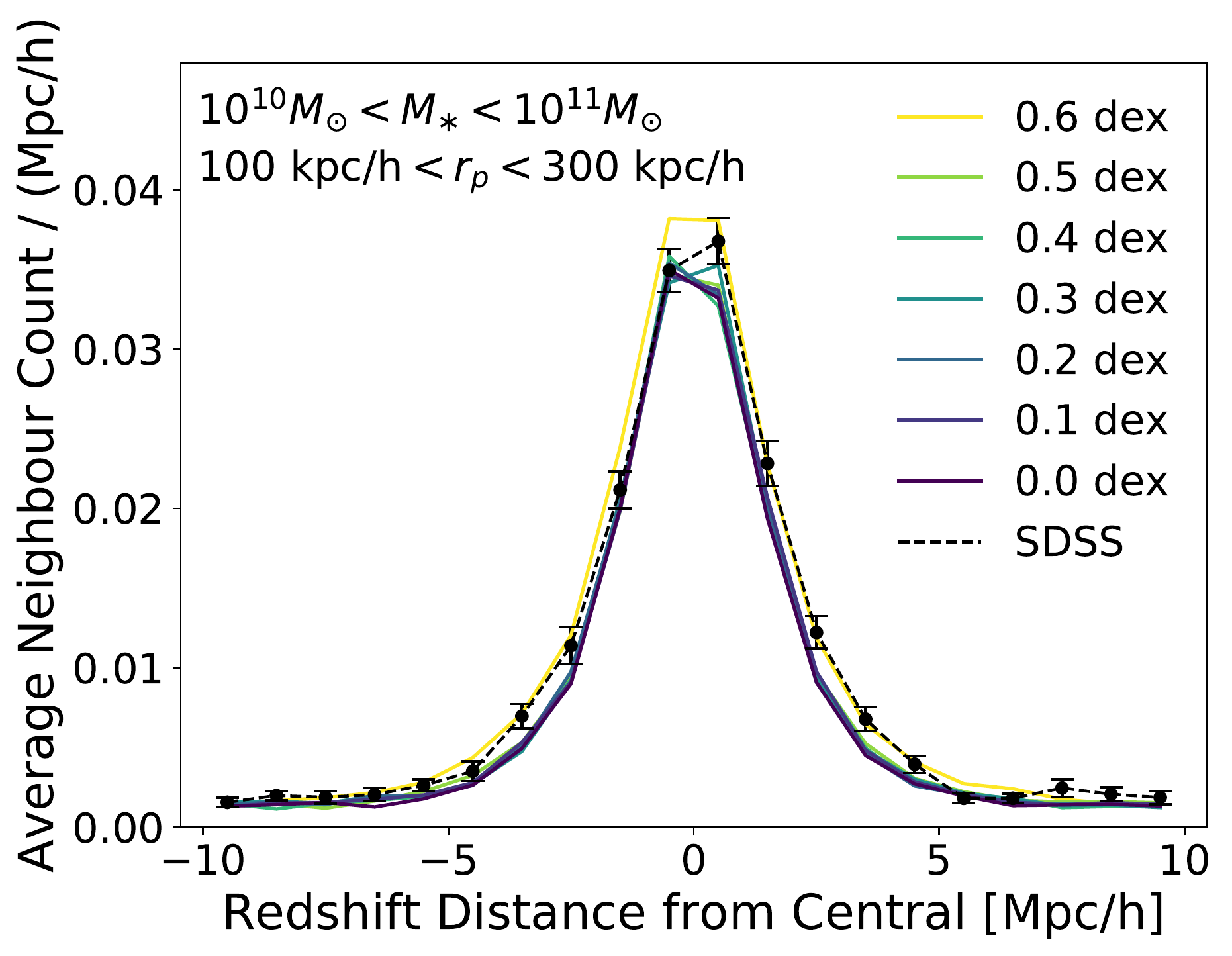}
    \caption{Comparison of observed and simulated velocity distribution functions for high-mass galaxies.  Neighbours are galaxies in the same mass range and within 100-300 kpc/h projected distance. }
    \label{fig:high_vdf}
\end{figure}

\begin{figure}
	\includegraphics[width=\columnwidth]{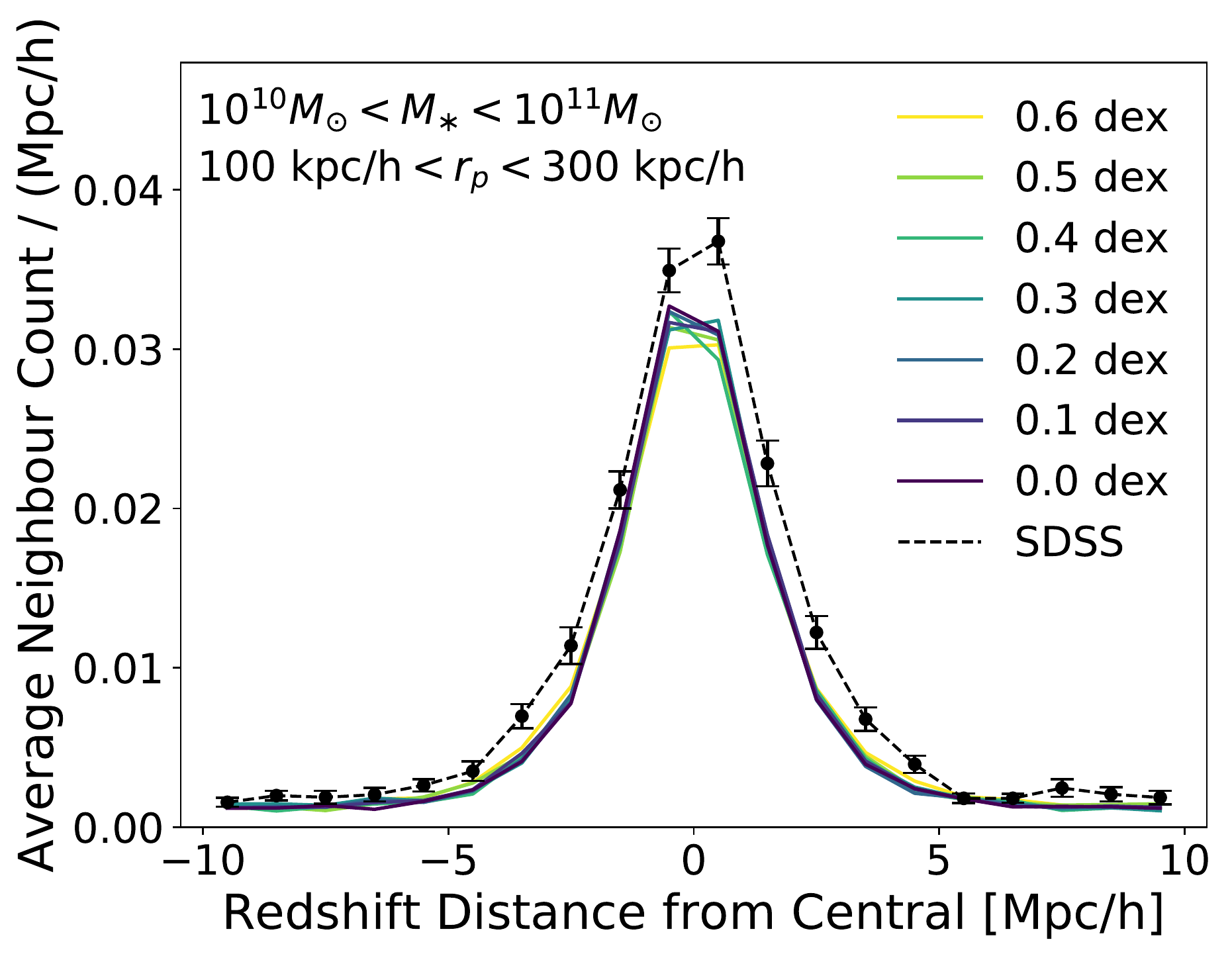}
    \caption{Comparison of observed and simulated velocity distribution functions for high-mass galaxies, with falsely identified centrals removed from the simulated distributions. Neighbours are galaxies in the same mass range and within 100-300 kpc/h projected distance. }
    \label{fig:high_vdf_UPID}
\end{figure}

Higher values of scatter allow larger haloes to host smaller galaxies and vice versa, and so, like the CSMFs, VDFs are expected to increase for low-mass central galaxies and decrease for high-mass central galaxies.  We find the expected trend clearly differentiated for low-mass central galaxies (Figure \ref{fig:low_vdf}). Again, there is little distinction in high-mass central galaxies, except for the outlying 0.6 dex VDF that does not conform to the expected trend (Figure \ref{fig:high_vdf}). This is the same effect seen in the CSMFs (\S \ref{sec:csmf}), where increased scatter causes contamination in the sample, reducing purity. With falsely identified centrals removed from the high-mass sample, the counts fall with increasing scatter, as expected (Figure \ref{fig:high_vdf_UPID}). 

Given that our simulated catalogues have no difference for the scatter for high-mass ($\gtrsim 10^{12}\Msun$) haloes, we expect (and find) that the greatest sensitivity to the input scatter comes from the CSMFs and VDFs for low-mass galaxies.  Because the CSMFs and VDFs for high-mass galaxies are relatively insensitive to the input scatter, they instead serve as an important check that our adopted orphan model is realistic---i.e., that our satellite lifetimes are accurate.


\section{Results and Discussion} \label{sec:results} \label{sec:discussion}

We compare the observed distributions to the simulations for varying values of scatter. No significant discrepancies exist between the models and the observations (Figures \ref{fig:low_csmf}, \ref{fig:low_vdf}, \ref{fig:high_csmf}, and \ref{fig:high_vdf}), giving confidence that the models are sufficiently flexible to describe the observations.

The CSMF and VDF for low-mass central galaxies are shown in Figures \ref{fig:low_csmf} and \ref{fig:low_vdf}, respectively. They indicate that the observations are most consistent with values of scatter $<$0.6 dex. The high-mass CSMF and VDF (Figures \ref{fig:high_csmf} and \ref{fig:high_vdf}) show less distinction across scatters, but are consistent with the simulated catalogues, suggesting that our orphan model is correctly capturing satellite lifetimes. Thus, we cannot put tight constraints on scatter without improved statistics beyond those available in the SDSS. 

The methods we developed in \S \ref{sec:methods} nonetheless have the potential power to distinguish between different scatters for low-mass haloes and produce distributions that are consistent with the observed Universe. This method could be applied in the future to larger data sets; in particular, the upcoming Dark Energy Spectroscopic Instrument (DESI) Bright Galaxy Survey is anticipated to provide redshifts for over 10 million galaxies \citep{DESI} compared to the $\sim$200,000 SDSS targets in our cut, and the pipeline will be able to distinguish close neighbor luminosity profiles with a higher degree of precision \citep{DESIpipeline}.  Even regardless of the ability to probe satellites at closer projected distances, the resulting error bars would be better by at least a factor of $\sqrt{50}$, enabling very sensitive tests for both the orphan model adopted and for the scatter in the stellar mass--halo mass relation for galaxies with $M_h<10^{11} \Msun$.  Based on the scatter-dependence of the low-mass CSMF (Figure \ref{fig:low_csmf}), such reduced error bars could constrain scatter to approximately $\pm$0.05 dex and allow sensitive testing of satellite lifetimes and the orphan model described in \S \ref{sec:sims}. 

\section{Conclusions} \label{sec:conclusions}

We have developed a method (\S \ref{sec:methods}) to constrain scatter in the stellar mass--halo mass relation for central galaxies with host halo masses $M_h\le 10^{11} \Msun$ by comparing the mass and velocity distributions of neighbouring galaxies in dark matter simulations and observations.  We find that the simulations produce very consistent results with the Sloan Digital Sky Survey for most values of scatter ($<$0.6 dex),  and that SDSS statistics are inadequate to firmly constrain the scatter (\S \ref{sec:discussion}). The method is nonetheless easily applied to future surveys, including the Dark Energy Spectroscopic Instrument experiment (DESI; \citealt{DESI}), which will conclusively measure the stellar mass--halo mass scatter for low-mass haloes. 



\section*{Acknowledgements}

PB was partially supported through program number HST-HF2-51353.001-A, provided by NASA through a Hubble Fellowship grant from the Space Telescope Science Institute, which is operated by the Association of Universities for Research in Astronomy, Incorporated, under NASA contract NAS5-26555.

\bibliographystyle{mnras}
{\footnotesize
\bibliography{master_bib}
}

\appendix

\section{Isolation cuts} \label{a:purity_completeness}

\begin{figure*}
    \centering
    \subfloat[Low-mass galaxies ($10^{9}$-$10^{10} \Msun$).]{
      \includegraphics[width=0.9\columnwidth]{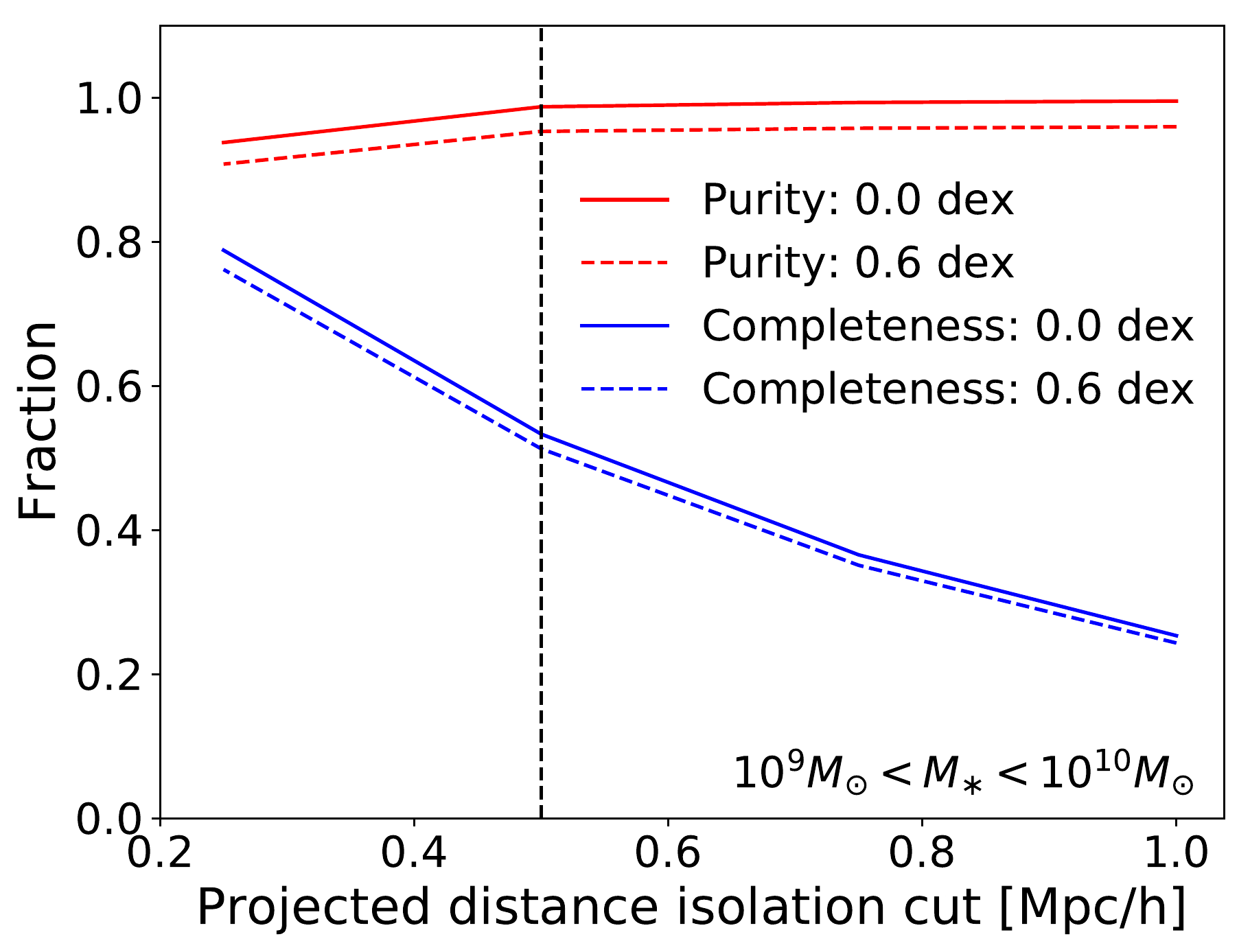}
    }\qquad
    \subfloat[High-mass galaxies ($10^{10}$-$10^{11} \Msun$).]{
     \includegraphics[width=0.9\columnwidth]{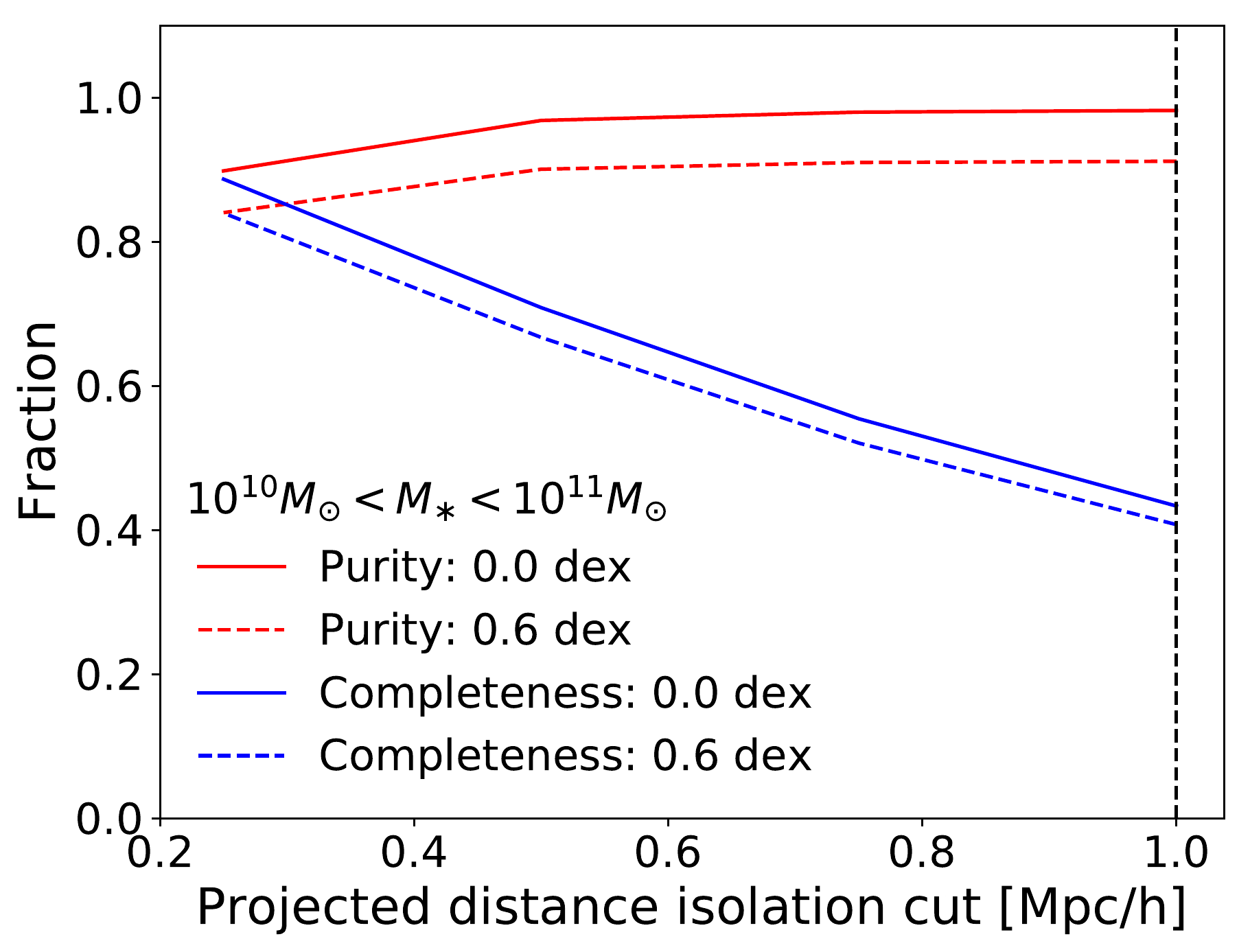}
    }
    \caption{Purity and completeness for the central--satellite selection criteria for low-mass (left) and high-mass (right) simulated galaxies as a function of projected distance cut. Purity shown in red and completeness in blue (solid for 0.0 dex scatter, dashed for 0.6 dex scatter); the dashed vertical line indicates our chosen centrality cut.}
    \label{fig:pc_by_cr}
\end{figure*}

\begin{figure*}
    \centering
    \subfloat[Low-mass galaxies ($10^{9}$-$10^{10} \Msun$).]{
      \includegraphics[width=0.9\columnwidth]{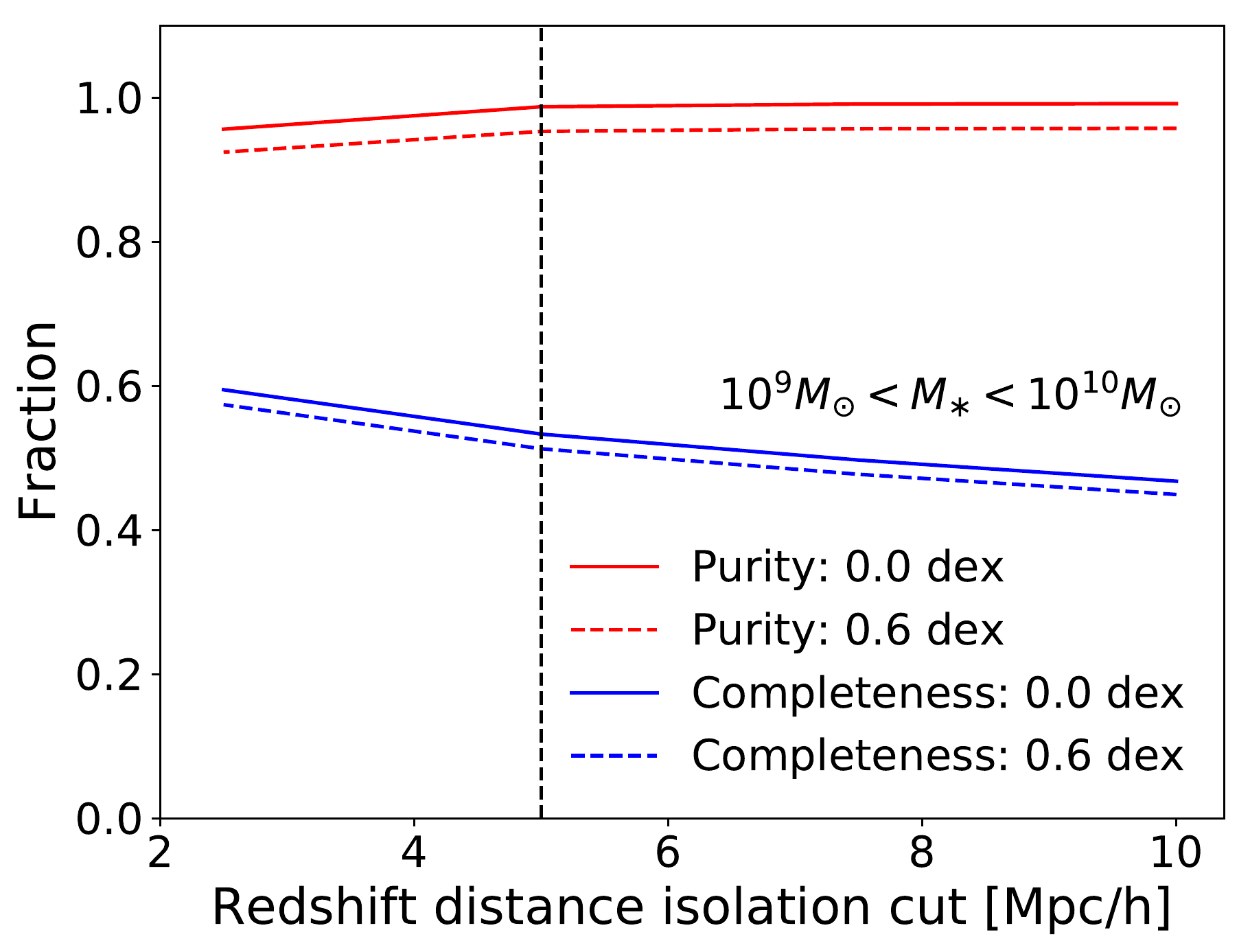}
    }\qquad
    \subfloat[High-mass galaxies ($10^{10}$-$10^{11} \Msun$).]{
     \includegraphics[width=0.9\columnwidth]{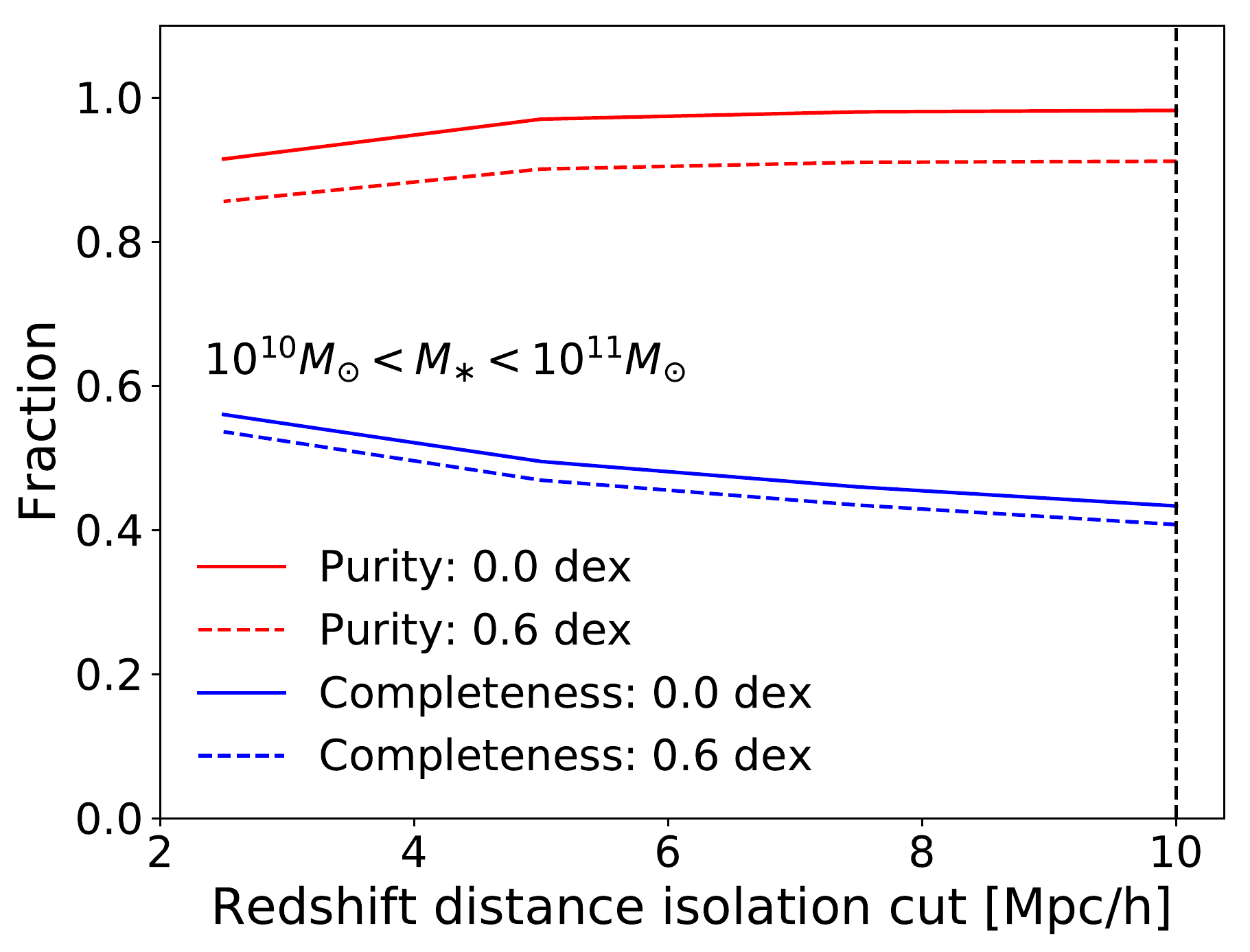}
    }
    \caption{Purity and completeness for the central--satellite selection criteria for low-mass (left) and high-mass (right) simulated galaxies as a function of redshift distance cut. Purity shown in red and completeness in blue (solid for 0.0 dex scatter, dashed for 0.6 dex scatter); the dashed vertical line indicates our chosen centrality cut.}
    \label{fig:pc_by_cz}
\end{figure*}

Isolation cuts are applied to galaxies in both simulations at observations to select central galaxies. We choose cuts for low-mass ($10^9$-$10^{10} M_{\odot}$) and high-mass ($10^{10}$-$10^{11} M_{\odot}$) galaxies to maximize purity and completeness, i.e., the ratio of simulated centrals matching our centrality cut to total centrality cut or total simulated centrals, respectively. We took cuts to maximize completeness while retaining above 90\% purity for all values of scatter; both decrease monotonically with increasing scatter, but vary by less than 10\% over 0.0-0.6 dex from typical values of $\sim$95\% purity and $\sim$50\% completeness (Table \ref{tab:purity_completeness} in \S \ref{sec:methods}). As expected, purity rises and completeness falls with harsher isolation cuts for both low-mass and high-mass galaxies (Figures \ref{fig:pc_by_cr} and \ref{fig:pc_by_cz}). 

\section{Orphan haloes in simulations} \label{a:orphans}


Orphan satellite haloes are often included in simulations to match galaxy clustering, as described in \S \ref{sec:sims}. After generating catalogues with and without orphans, we find that models including orphan haloes provide a better match to observations.

\subsection{Inclusion of orphan haloes}  \label{a:include_orphans}

\begin{figure*}
    \centering
    \subfloat[Low-mass galaxies ($10^{9}$-$10^{10} \Msun$).]{
      \includegraphics[width=0.9\columnwidth]{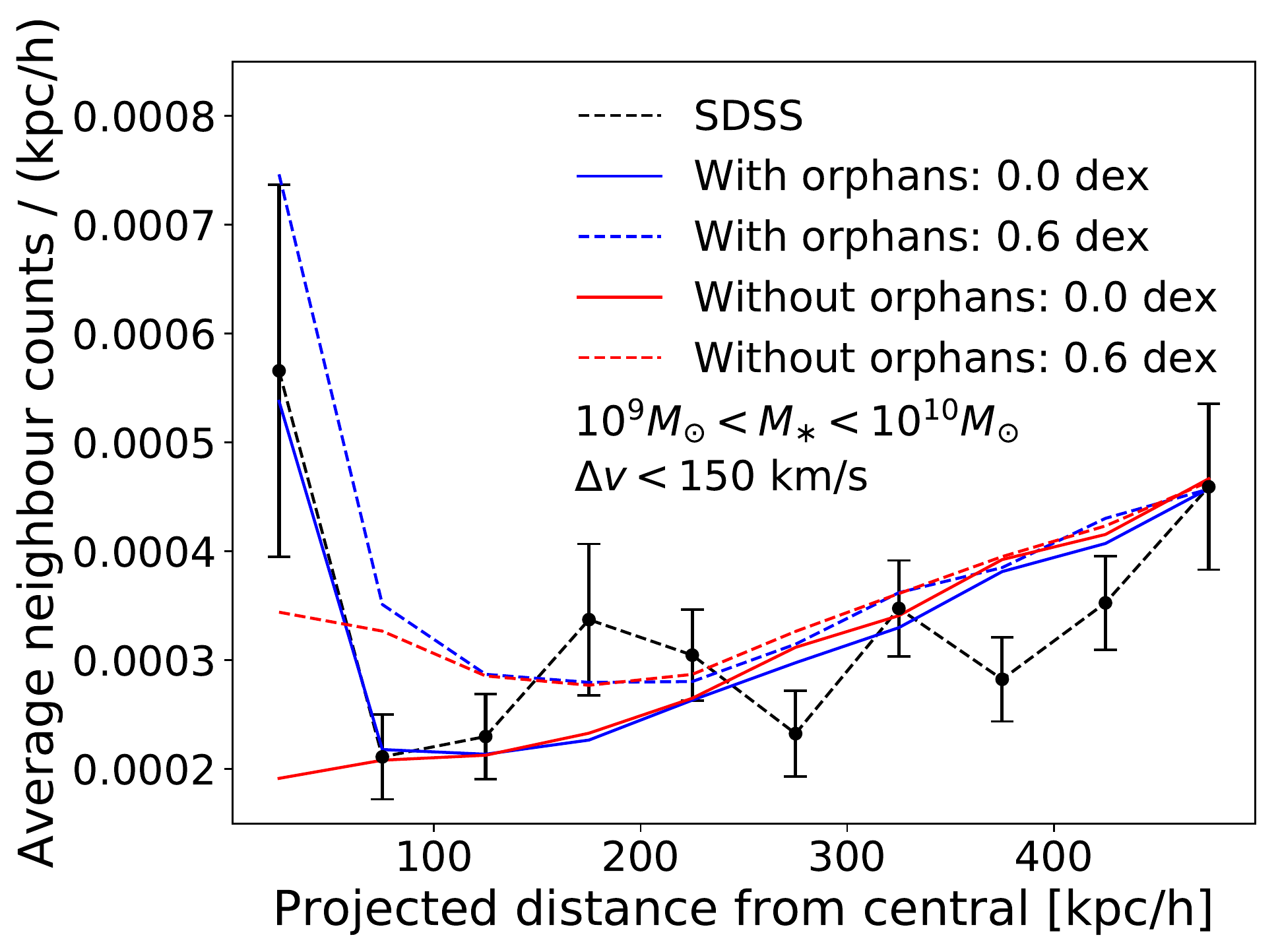}
    }\qquad
    \subfloat[High-mass galaxies ($10^{10}$-$10^{11} \Msun$).]{
     \includegraphics[width=0.9\columnwidth]{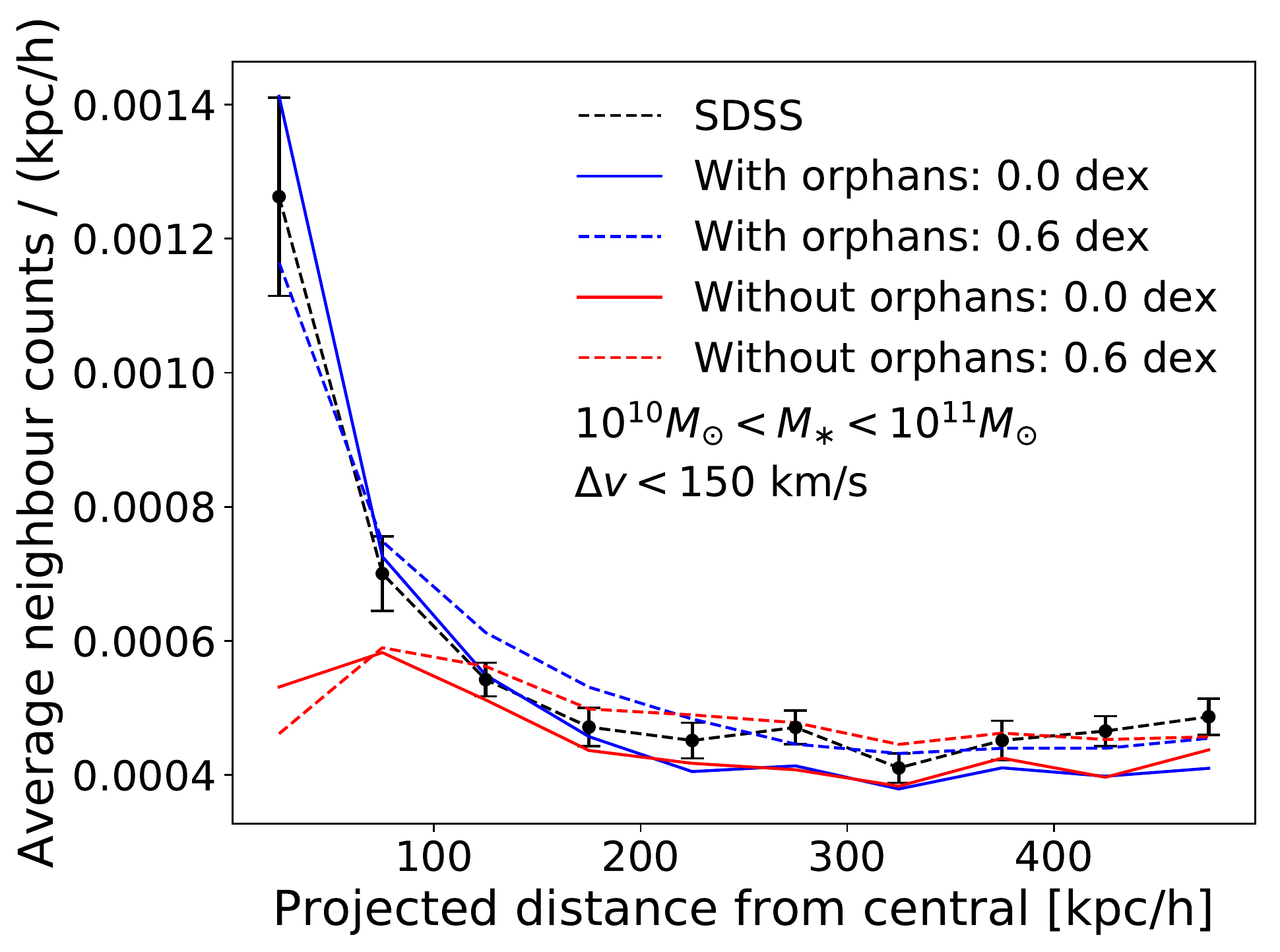}
    }
    \caption{Average neighbour counts as a function of projected distance to low-mass (left) and high-mass (right) central galaxies, for catalogues generated with and without orphans. Neighbours are within $\pm 1.5$ Mpc/h redshift distance. The black line shows the observed distribution, compared to the simulation with and without orphan haloes (blue and red lines, respectively) for different scatters.}
    \label{fig:nbyr_compare_orphans}
\end{figure*}

Average neighbour counts as a function of projected distance from their central galaxy show agreement with the catalogue including orphans, for both low- and high-mass galaxies (Figure \ref{fig:nbyr_compare_orphans}). Errors on observations are calculated with the bootstrap resampling method described in \S \ref{sec:obs_err}. Clearly, catalogues generated without orphans grossly under-represent neighbours within 150 kpc/h, while catalogues with orphans place these neighbour counts within a sigma for high and low scatters. 

\subsection{Motivating projected distance cuts}  \label{a:rp_cuts}

The characteristics of a galaxy's nearest larger neighbour provide valuable information about clustering and serves as a measure of consistency between simulated and observed catalogues, motivating projected distance cuts on neighbours in the CSMFs and VDFs. The analysis below is carried out for catalogues adopting zero scatter, and the results are similar for 0.6 dex. 

\begin{figure}
	\includegraphics[width=\columnwidth]{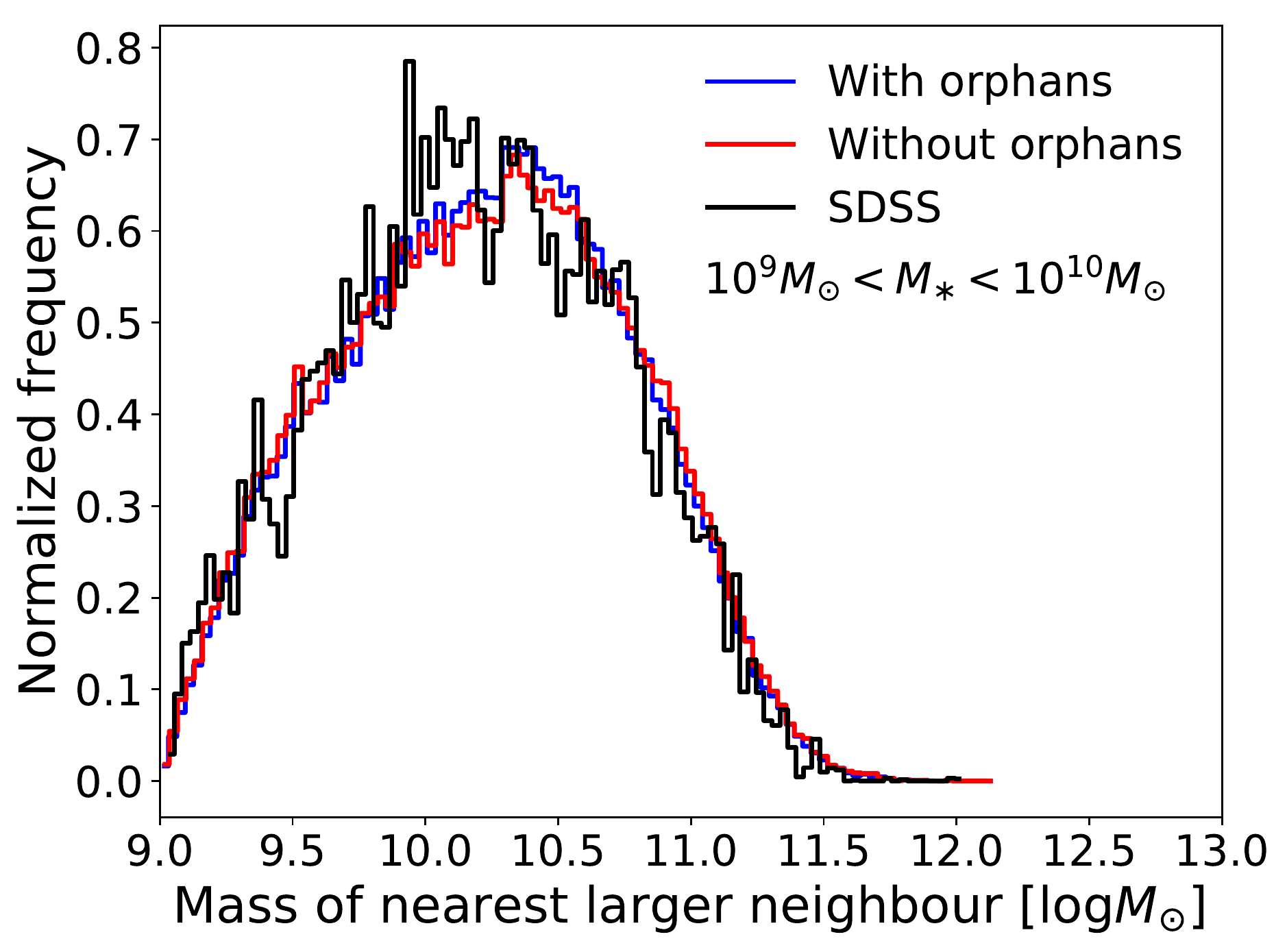}
    \caption{Normalized histogram of mass of nearest larger neighbour for high-mass galaxies (neighbours within 0.5 Mpc/h projected and $\pm 5$ Mpc/h redshift distance). Compares observations (black), catalogue with orphans for (blue), and catalogue without orphans (red). Simulations are both for the model with zero scatter.}
    \label{fig:low_nln_mass_hist}
\end{figure}

\begin{figure}
	\includegraphics[width=\columnwidth]{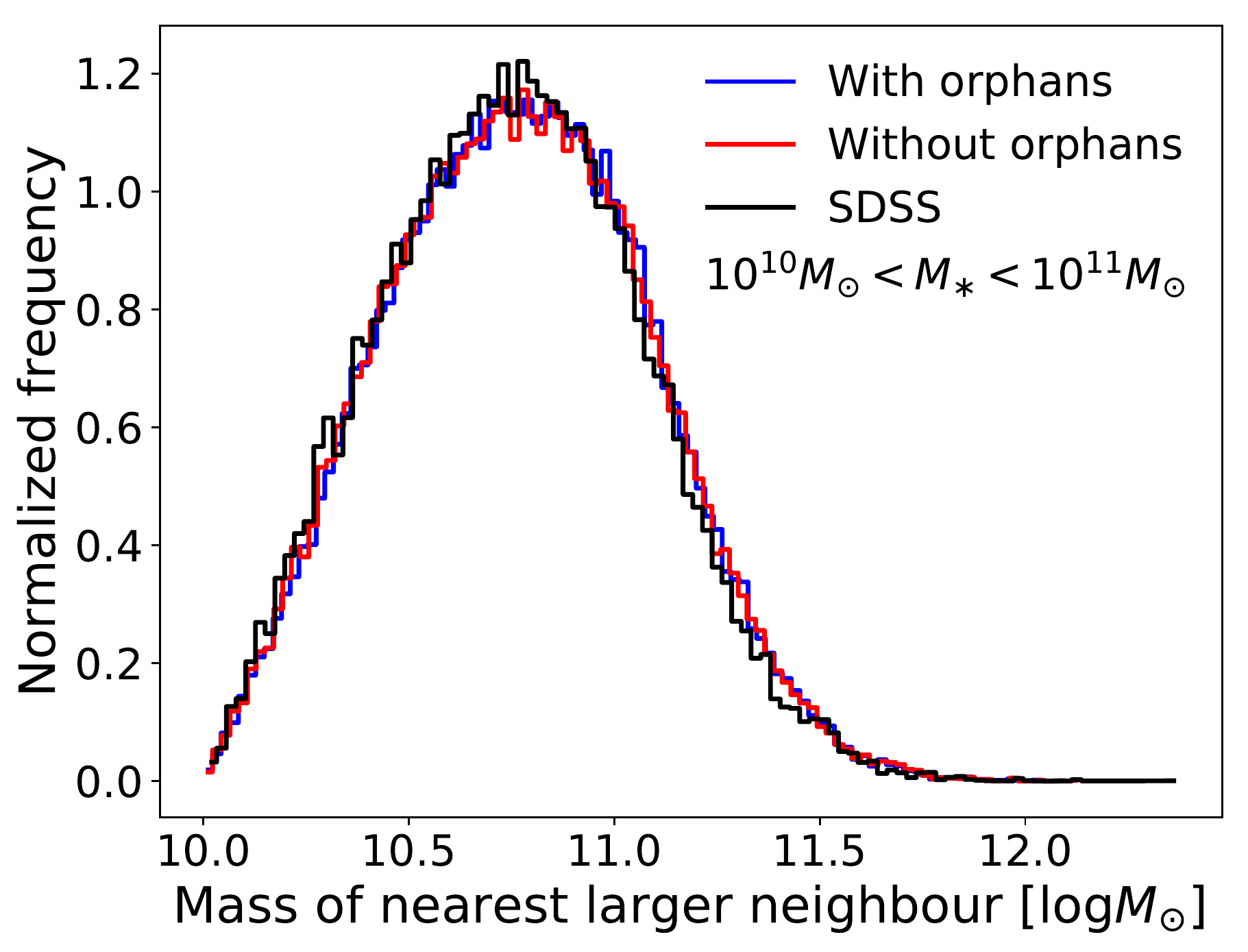}
    \caption{Normalized histogram of mass of nearest larger neighbour for high-mass galaxies (neighbours within 1 Mpc/h projected and $\pm 10$ Mpc/h redshift distance). Compares observations (black), catalogue with orphans for (blue), and catalogue without orphans (red). Simulations are both for the model with zero scatter.}
    \label{fig:high_nln_mass_hist}
\end{figure}

The mass distributions of nearest larger neighbours show close agreement between observations and both catalogues, for low- and high-mass galaxies (Figures \ref{fig:low_nln_mass_hist} and \ref{fig:high_nln_mass_hist}). 


\begin{figure}
	\includegraphics[width=\columnwidth]{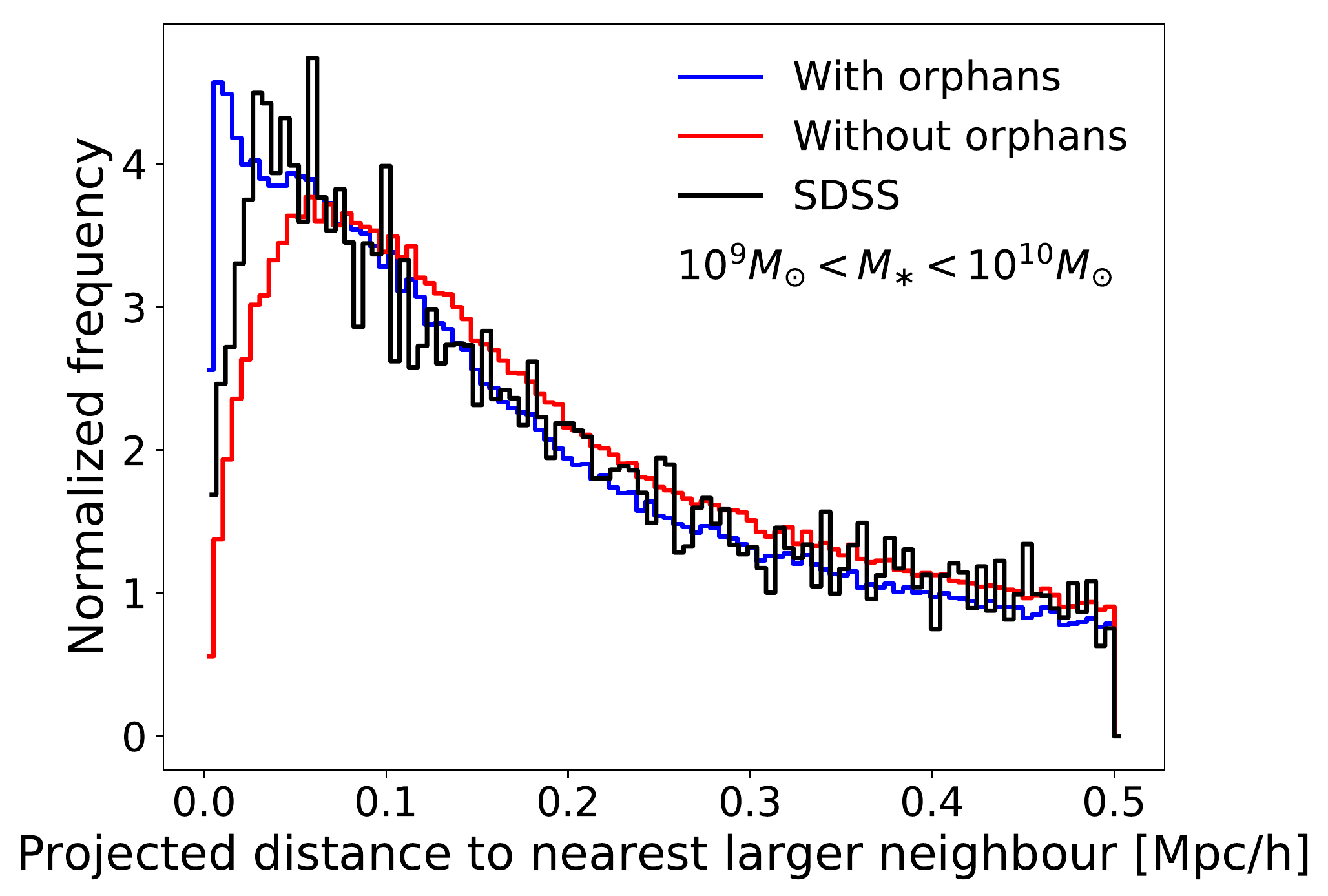}
    \caption{Normalized histogram of projected distance to nearest larger neighbour for high-mass galaxies (neighbours within 0.5 Mpc/h projected and $\pm 5$ Mpc/h redshift distance). Compares observations (black), catalogue with orphans for (blue), and catalogue without orphans (red). Simulations are both for the model with zero scatter.}
    \label{fig:low_nln_projdist_hist}
\end{figure}

\begin{figure}
	\includegraphics[width=\columnwidth]{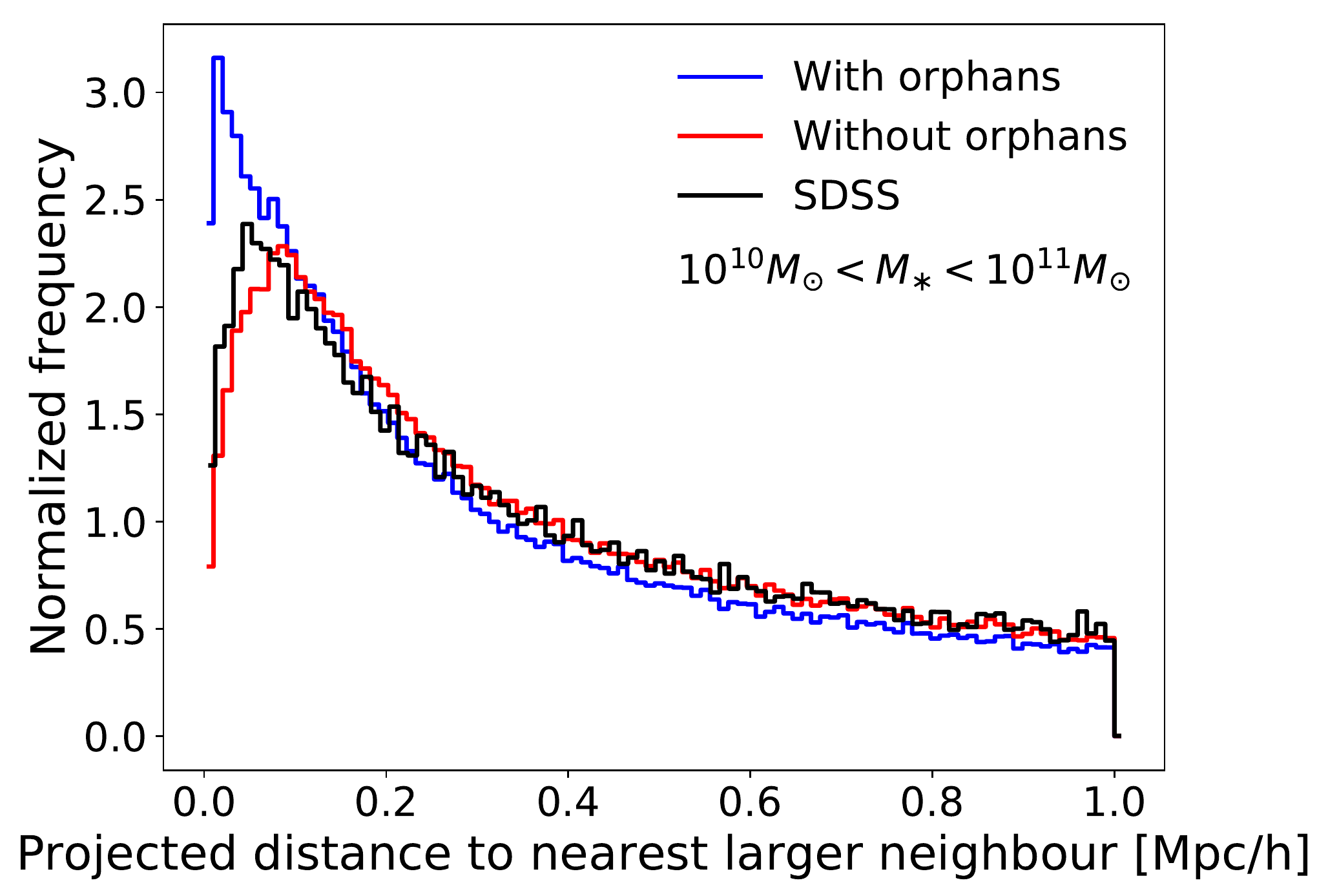}
    \caption{Normalized histogram of projected distance to nearest larger neighbour for high-mass galaxies (neighbours within 1 Mpc/h projected and $\pm 10$ Mpc/h redshift distance). Compares observations (black), catalogue with orphans for (blue), and catalogue without orphans (red). Simulations are both for the model with zero scatter.}
    \label{fig:high_nln_projdist_hist}
\end{figure}

Notable differences are seen in the distribution of projected distance to nearest larger neighbour for very near ($<$100 kpc/h) neighbours, in the case of both low- and high-mass galaxies (Figures \ref{fig:low_nln_projdist_hist} and \ref{fig:high_nln_projdist_hist}). Misleadingly, the high-mass distribution appears to more closely match catalogues without orphans.


This apparent deficit in near neighbours, particularly for high-mass galaxies, is likely due to observational incompleteness, i.e., overlapping luminosity profiles resulting in near neighbours going undetected. This effect would be more prominent in high-mass galaxies, since they have brighter luminosities. Inaccuracy in fiber collision corrections (described in \S \ref{sec:obs_corr}) would also more greatly effect high-mass galaxies, which have higher degrees of spatial clustering. Galaxies this close may also be in the process of merging, which would further complicate disentangling them in the observational pipelines.

\begin{figure}
	\includegraphics[width=\columnwidth]{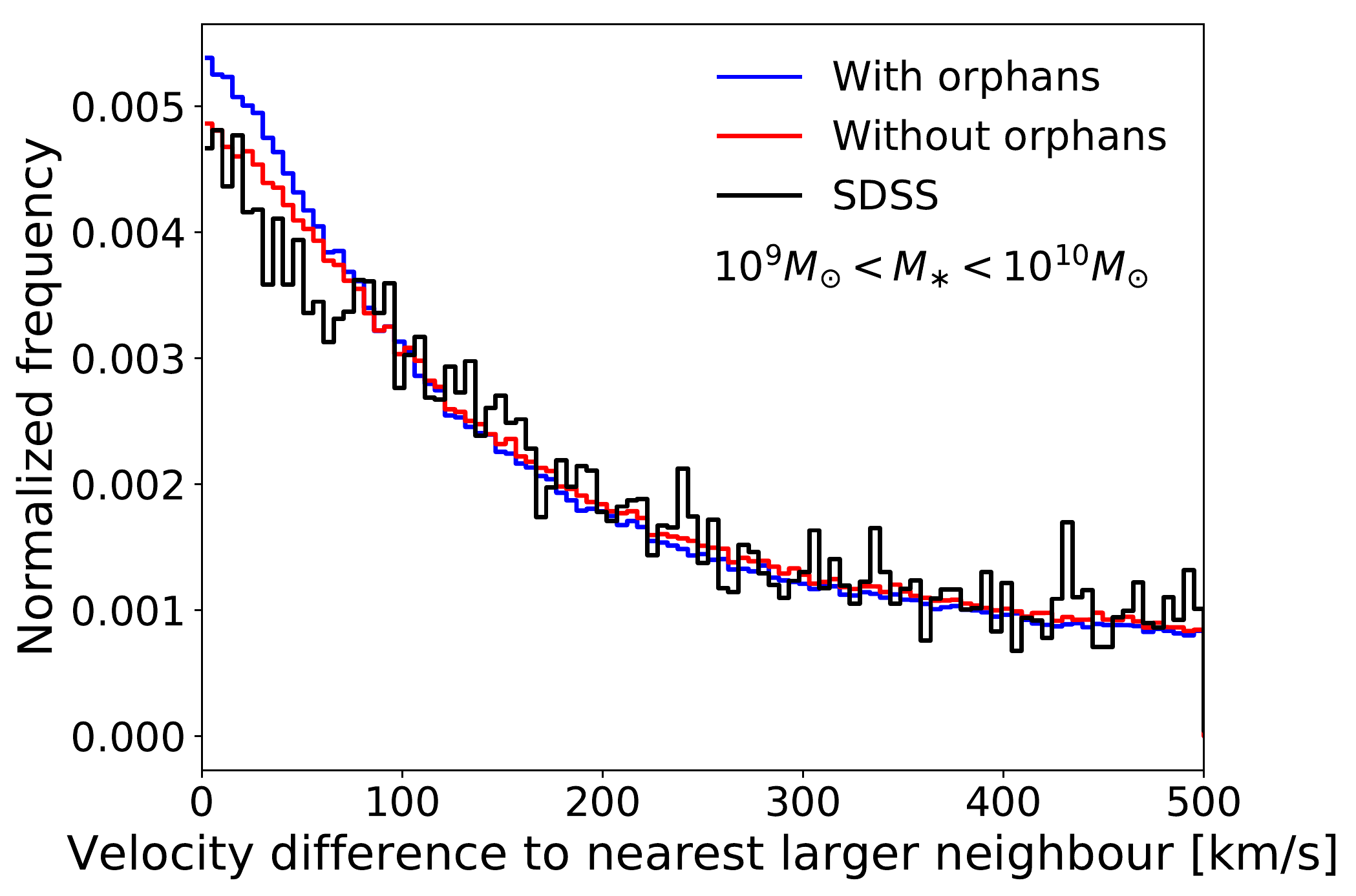}
    \caption{Normalized histogram of magnitude of velocity difference to nearest larger neighbour for high-mass galaxies (neighbours within 0.5 Mpc/h projected and $\pm 5$ Mpc/h redshift distance). Compares observations (black), catalogue with orphans for (blue), and catalogue without orphans (red). Simulations are both for the model with zero scatter.}
    \label{fig:low_nln_veldiff_hist}
\end{figure}

\begin{figure}
	\includegraphics[width=\columnwidth]{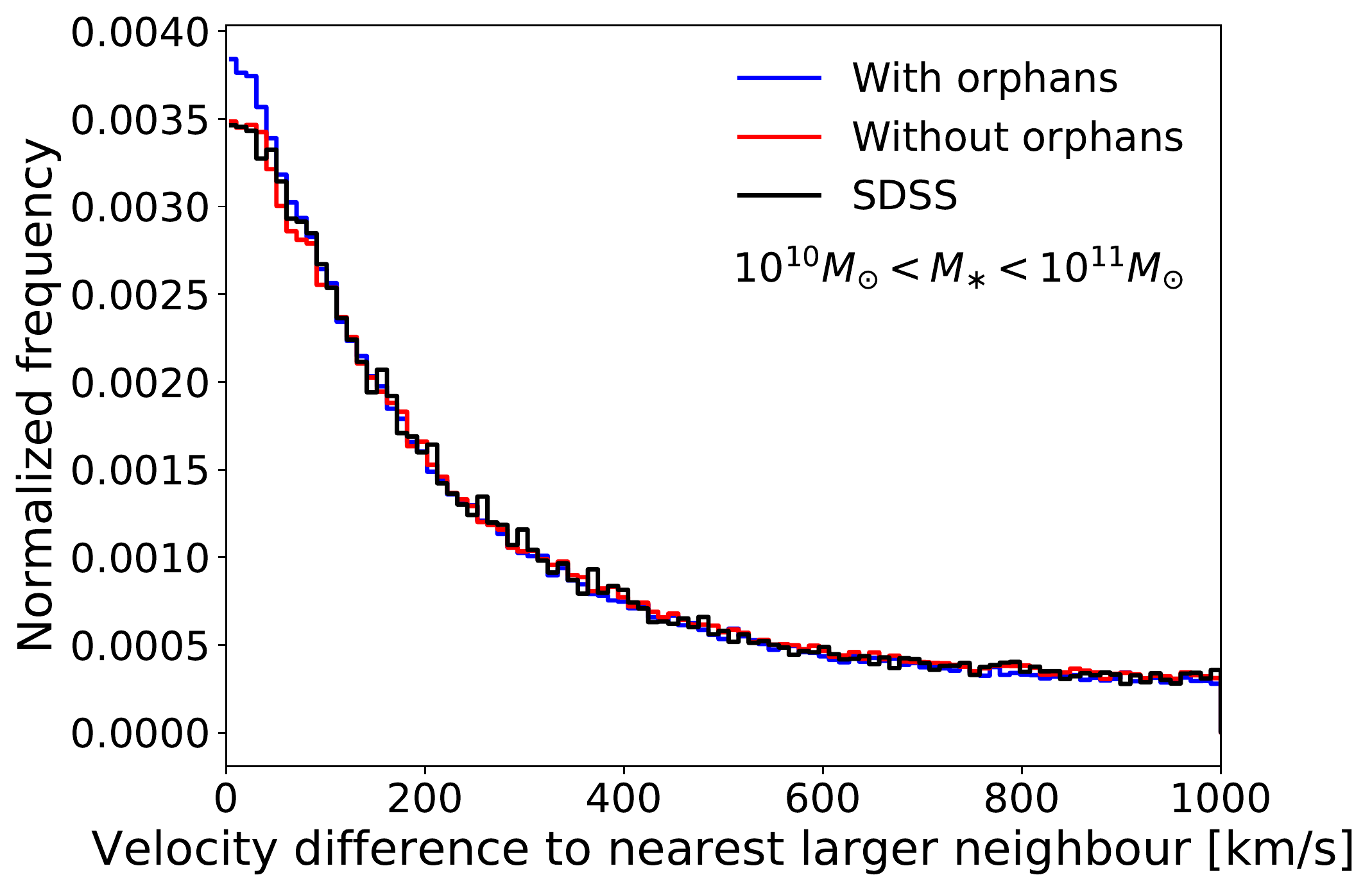}
    \caption{Normalized histogram of magnitude of velocity difference to nearest larger neighbour for high-mass galaxies (neighbours within 1 Mpc/h projected and $\pm 10$ Mpc/h redshift distance). Compares observations (black), catalogue with orphans for (blue), and catalogue without orphans (red). Simulations are both for the model with zero scatter.}
    \label{fig:high_nln_veldiff_hist}
\end{figure}

The distribution of velocity differences to nearest larger neighbours is shown in Figures \ref{fig:low_nln_veldiff_hist} and \ref{fig:high_nln_veldiff_hist}; a normal distribution of noise centred on $\pm$30 km/s is added to the simulated velocity distributions to match uncertainties in redshift measurements. Observations are similar to both catalogues, but show slight differences from the catalogue with orphans at very close ($<$50 km/s) redshift distances. Neighbours with small line-of-sight velocity differences may correspond to small projected distances as well, and then suffer the same issues of completeness as described above. 


Overall, the differences in the distributions of nearest larger neighbour characteristics between observations and simulated catalogues (with or without orphan haloes) are small, and we use these results to motivate a projected distance cut of $>$50 kpc/h and $>$100 kpc/h for the low- and high-mass CSMFs and VDFs described in \S \ref{sec:csmf} and \S \ref{sec:v_dist}, respectively, and thus minimize these effects in our analysis.

\bsp	
\label{lastpage}
\end{document}